\documentclass[]{aastex631}

\usepackage{chemformula}
\usepackage{upgreek}
\usepackage{mathtools}

\submitjournal{The Planetary Science Journal} 

\shorttitle{Sputter Models}
\shortauthors{Jäggi et al.}

\renewcommand{\edit}[2]{#2} 

\begin{document}

\title{New compound and hybrid binding energy sputter model for modeling purposes\\ in agreement with experimental data}


\correspondingauthor{Noah Jäggi}
\email{noah.jaeggi@unibe.ch}

\author[0000-0002-2740-7965]{Noah Jäggi}
\affil{Physikalisches Institut, University of Bern, Sidlerstrasse 5, CH-3012 Bern, Switzerland}

\author{Andreas Mutzke}
\affil{Max Planck Institute for Plasma Physics (IPP), D-17491 Greifswald, Germany}

\author[0000-0002-9854-2056]{Herbert Biber}

\author[0000-0001-9999-9528]{Johannes Brötzner}

\affil{Institute of Applied Physics, TU Wien, Wiedner Hauptstraße 8-10/E134, A-1040 Vienna, Austria}

\author[0000-0002-7478-7999]{Paul Stefan Szabo}

\affil{Space Sciences Laboratory, University of California, 7 Gauss Way, Berkeley, 94720 CA, USA}

\author[0000-0002-9788-0934]{Friedrich Aumayr}

\affil{Institute of Applied Physics, TU Wien, Wiedner Hauptstraße 8-10/E134, A-1040 Vienna, Austria}

\author[0000-0002-2603-1169]{Peter Wurz}
\author[0000-0003-2425-3793]{André Galli}
\affil{Physikalisches Institut, University of Bern, Sidlerstrasse 5, CH-3012 Bern, Switzerland}

\begin{abstract}
Rocky planets and moons experiencing solar wind sputtering are continuously supplying their enveloping exosphere with ejected neutral atoms. To understand the quantity and properties of the ejecta, well established Binary Collision Approximation Monte Carlo codes like TRIM with default settings are used predominantly. Improved models such as SDTrimSP have come forward and together with new experimental data the underlying assumptions have been challenged. We introduce a hybrid model, combining the previous surface binding approach with a new bulk binding model akin to \cite{Hofsass2022}. In addition, we expand the model implementation by distinguishing between free and bound components sourced from mineral compounds such as oxides or sulfides. The use of oxides and sulfides also enables the correct setting of the mass densities of minerals, which was previously limited to the manual setting of individual atomic densities of elements. All of the energies and densities used are thereby based on tabulated data, so that only minimal user input and no fitting of parameters are required. We found unprecedented agreement between the newly implemented hybrid model and previously published sputter yields for incidence angles up to 45$^\circ$ from surface normal. Good agreement is found for the angular distribution of mass sputtered from enstatite \ch{MgSiO3} compared to latest experimental data. Energy distributions recreate trends of experimental data of oxidized metals. Similar trends are to be expected from future mineral experimental data. The model thus serves its purpose of widespread applicability and ease of use for modelers of rocky body exospheres.
\end{abstract}

\section{Introduction}\label{sec:intro}
In recent years there were several efforts to better constrain the erosion of rocky planetary bodies exposed to highly energetic solar wind ions. This includes investigating the effect of surface roughness \citep{Biber2022} and porosity \citep{Szabo2022b}, performing ion irradiation experiments with mass yield measurements \citep[e.g.,][]{Hijazi2017,Szabo2018,Szabo2020a,Biber2022} as well as new surface- and bulk-binding-energy  model from theory \citep[][]{Hofsass2022,Morrissey2022}. In this work, we discuss the parameter of density and its inclusion in SDTrimSP \citep{Mutzke2019} as well as a new hybrid binding energy model that reliably recreates experimental sputter yields completely without the requirement to adjust input parameters. The new approach will pose a valuable tool for modeling the ion sputtering contribution to exospheres \citep[i.e.,][]{Pfleger2015,Suzuki2020,Killen2022,Kazakov2022}.

\subsection{Space weathering of exposed rocky surfaces}
Exposed bodies in space are subject to solar wind irradiation. The main constituents of solar wind, \ch{H+}, and \ch{He^2+}, thereby bear kinetic energies of approximately 1~keV/amu---equivalent to about 440~km/s \citep{Wurz2005,Gershman2012,Winslow2013,Baker2013}. When hitting a surface, most ions are neutralized and enter the sample, with some fraction being reflected as either neutrals or even ions \citep{Lue2011,Vorburger2013}. The ions entering the sample initiate a cascade of collisions with a chance to eject particles from the near-surface at supra-thermal energies. This process is responsible for altering the surface composition and creating lattice defects which leads to amorphization \citep[][]{Betz1994,Loeffler2009,Dukes2011,Domingue2014}. 

Ion sputtering releases atoms from the surface having typical velocities that are significantly lower than the impinging ions \citep[e.g.][]{Thompson1968}, but large enough to form an extended exosphere with a significant fraction of atoms exceeding the escape velocity of any small body including the Moon (2.4~km/s) and Mercury (4.3~km/s) \citep[e.g.,][]{Wurz2007,Wurz2010}. Such exospheres allow for ground-based observatories and space probe missions such as LADEE and LRO at the Moon \citep{Paige2010,Elphic2014} and MESSENGER \citep{Solomon2001,McNuttJr2018} or the future BepiColombo \citep{Benkhoff2010,Milillo2020,Orsini2020} at Mercury to detect them.  
These observations were used early on to self-consistently model Mercury's surface composition based on the four expected major processes contributing to the exosphere: Solar wind ion sputtering, micro-meteroid impact vaporization, photon-stimulated desorption, and thermal desorption \citep{Madey2002,Mura2009,Wurz2010,Gamborino2018,Wurz2022}. 

An important piece of information which is necessary to distinguish the exospheric species sourced from the surface is the process-specific energy distribution of the ejected material. For example, solar wind ion sputtering and micro-meteroid impact vaporization compete in supplying Mercury’s exospheric high-energy particle population with refractory species (e.g., Ca and Mg), whilst photon-stimulated desorption dominates the supply of energetic volatile and moderately volatile species (i.e., Na, K, and S) \cite{Mangano2007,Cassidy2015,Schaible2020,Janches2021,Grava2021}. The same way that fluxes, or precipitation rates, of the particles causing these processes are still in the process of being better constrained \citep[i.e., proton precipitation for solar wind sputtering at Mercury's cusps in][]{Fatemi2020,Raines2022,Glass2022}, the understanding of the underlying physics is still a work in progress. At the Moon, precipitation rates seem comparably trivial to compute, but the Moon travelling through the Earth's magnetotail as well as localized crustal fields add complexity to the system \citep[e.g.,][]{Lue2011,Poppe2018,Nenon2020}. 

\subsection{Sputter models}
To efficiently model ion induced sputtering, Binary Collision Approximation (BCA) models are used. The BCA codes track particles as they travel through the sample and cause recoils, which are in turn tracked throughout the sample. There are many different models available, however, we will focus on the results of the Monte Carlo based, most widely used TRIM code \citep{Biersack1980} in the SRIM package \citep{Ziegler2010} as well as its successor SDTrimSP \citep{Mutzke2019}, a combined and improved version of the static TRIM.SP \citep{Biersack1984} and the dynamic TRYDIN \citep{Moller1984}.

TRIM has been shown to overestimate the sputter yield compared to experimental yields for minerals \citep{Szabo2018}. Exosphere modelers need more accurate inputs which are in line with the latest understanding of sputtering. There have been several suggestions on how to best recreate experimental data. Here are the major contributions that set the expectations as well as limitations of the current state-of-the-art sputter modeling.

\begin{itemize}
    \item \cite{Schaible2017} varied O binding energies to better fit early experimental data for sputtering of \ch{Al2O3} and \ch{SiO2} \citep{Roth1979,KenKnight1967}. Increasing the O-binding energy decreases the O yield, but not enough to significantly improve the agreement.
    \item \cite{Szabo2020a} suggested that the best agreement between the mass yield of an irradiated sample and SDTrimSP is obtained by a) adjusting atomic densities to obtain an appropriate sample density, b) adjusting the surface binding energy (SBE) of O to 6.5~eV and c) set the SBEs of each element to the averaged SBE of all elements in the sample, resulting in a SBE which is highly dependent on the O concentration in the sample (Appendix~\ref{app:isbv2}). Although we found these parameters to work reasonably well for all kinds of silicates, the universality of these modifications is questionable.
    \item \cite{Morrissey2022} determined surface binding energies using molecular dynamics and suggests lower sputter yield across all surface species due to an increase in single component's binding energies. However, the restricted availability of species-specific surface binding energies prevents the applicability of the results on a broad range of minerals. This is also caused by the limited availability of interatomic potentials for each mineral system of interest.
    \item \cite{Hofsass2022} proposed completely neglecting surface binding energies and instead using only bulk binding energies from tabulated data. This way, particles leaving the sample do not have to overcome a surface potential and instead lose energy with each recoil. Although they solely use tabulated data to set the bulk binding energy and propose a sound physical constraint on the cutoff energy for the tracing of the particles, they are still required to make use of an undisclosed level of implantation to find good agreement with experimental data. 
    \item  \cite{Biber2022} used an in-house built ray-tracing code SPRAY \citep{Cupak2021} with data from SDTrimSP and atomic force microscope images to discuss the effect of surface roughness on the sputter yield of a powder pellet and a flat, glassy thin film. They found that a rough pressed pellet surface reduces the yield, especially at shallow incident angles (above 45$^\circ$ relative to surface normal). The cause of this reduced yield was related to surface roughness leading to shallower local incident angles, shadowing, and re-deposition of material. For a detailed overview of rough surface sputter models see \cite{Kustner1998} and \cite{Arredondo2019}. 
\end{itemize}

All these models require varying degrees of adjustments of parameters when it comes to density, binding energies, cut-off energies or roughness. To adequately describe the sputtering process on realistic surfaces, roughness has to be taken into account. This effect is not considered in this work, as we focus on the fundamental sputter physics within the sample, which is agnostic to properties affecting trajectories of impinging ions and ejecta. For this reason, we compare our results to experimental thin-film data, which are considered to be flat surfaces \cite{Biber2022}. We propose a new compound model for obtaining a realistic initial mineral density as well as a hybrid binding energy model to obtain increased binding energies based on tabulated data that can recreate experimental results.


\section{Methods of Computation} \label{sec:methods}

\subsection{Model parameters}\label{met:SDTrimSP}
Angular dependent sputter yields for various different models were calculated with SDTrimSP to compare with a wide range of experimental data. To obtain good statistics in SDTrimSP, we modeled between $7.7\times10^6$ and $31\times10^6$ impactors for each of 19 incident angles between 0$^\circ$ and 89$^\circ$ relative to the surface normal \citep{Mutzke2019}. The step size was set to gradually decrease from an initial 10$^\circ$ for incidence close to the surface normal and dropping to 2$^\circ$ for incidence angles 80--88$^\circ$. We collected the information of up to $10^6$ recoils leaving the sample and perform statistics based on the last $10^5$ recoils. The data contains the species name, end energy, azimuth angle, and zenith angle. The fits of the data shown in the figures throughout this manuscript are described in Sec.~\ref{sec:fitting}. The inelastic loss model seven (inel~=~7) is used in all SDTrimSP calculations, which determines the inelastic loss in the sample based on the Lindhard-Scharff stopping power model \citep{Lindhard1961} unless there are corrections available \citep[e.g., tables for H and He in][]{Ziegler1985}. For a detailed description of SDTrimSP, we encourage the reader to look into the accompanying literature \citep[e.g.,][]{Mutzke2019}. 

The surface composition of irradiated samples show a clear fluence dependence until an equilibrium is reached. This was shown by \cite{Baretzky1992} for the oxide \ch{Ta2O5}, and by \cite{Szabo2020b} in the form of the fluence-dependence of experimental sputter data of minerals. Furthermore, the experimental sputter yields were best recreated using the dynamic mode of SDTrimSP \citep{Szabo2020b}. For this reason, all computations in this manuscript were performed in dynamic mode of SDTrimSP and the results are for ejecta from a surface in equilibrium with the impinging ions. For irradiation with He, the fluence was set to 750~at./\AA$^3$ whereas H irradiation required fluences of up to 3000~at./\AA$^3$ (or $3\times10^{19}$~at./cm$^3$) at normal incidence in some models. The dynamic mode allows the sample to change with the ion fluence and best simulates the sample composition reaching an equilibrium with the solar wind ions, reproducing the fluence-dependence of the experimental sputter yields. In detail, samples in SDTrimSP have an infinite lateral extent with a finite number of layers vertically. In our case, all layers have the same composition set initially and a thickness of 10~\AA. After each fluence step, comprised of about $10^5$ impactors, the layers within the sample are updated according to the components that were either lost or gained within the last step.

Direct comparisons between SRIM and SDTrimSP calculations were performed for mass yield (amu ion$^{-1}$). In SRIM \citep{Ziegler2010} we modeled $10^5$ impinging H and He ions for static sputter yield results to obtain good statistics. We used the `Monolayer Collision Steps / Surface Sputtering' damage model. The mineral density was set to its default density, as calculated by SRIM from the element components atomic density parameters (comparable to $\rho_\mathrm{atomic}$ from tabulated data in SDTrimSP given in Table~\ref{tab:major_minerals}).

We will now introduce a few select parameter settings that are required to model sputtering of minerals. These comprise of the dynamic mode of SDTrimSP, the different ways of introducing binding energies, including our new addition, as well as a new way for correcting sample density.

\subsection{Binding energy}\label{met:existing_binding_models}
The efficiency at which particles can be removed from a surface, the sputter yield, is in one part a function of the total binding energy of the system. The two common binding energies provided to a BCA model are the surface binding energy (SBE) and the bulk binding energy (BBE). The former is in the shape of a surface potential that has to be overcome to leave the sample. The latter is an energy that is subtracted from each recoil and simulates the interaction between neighboring atoms in the otherwise mineral-lattice-agnostic model that is SDTrimSP. It is possible to obtain a constant yield whilst keeping the sum of the binding energies constant \citep{Moller2001}. We now quickly introduce three different binding energy models, two of which are already established (pure SBE or BBE models) and one model that combines the two (SBE + BBE). The models are summarized in Table~\ref{tab:model_summary}.

\subsubsection{SB: Surface binding model}\label{met:SB_model}
The surface binding (SB) model is the default calculation model for TRIM and SDTrimSP. In this approach, a particle may leave the sample only if its kinetic energy exceeds the SBE. Energy loss within the sample occurs through elastic energy transfer during collisions and inelastic electronic losses.

Although the SBE is an energy determined by the attractive forces of neighboring atoms \citep{Sigmund1969,Gades1992}, it is common practice to approximate the SBE as the atomic enthalpy of sublimation ($\Delta H_S$). The exception are gasses where the surface binding energies are based on the enthalpy of dissociation. For example, pure O does not form a solid, and therefore the dissociation enthalpy of oxygen $\Delta H_{diss}$(\ch{O2}) is used instead of the sublimation enthalpy. Hobler and Morrissey showed for Si and Na that the atomic enthalpy of sublimation can severely underestimate the energy necessary to remove an atom from their crystalline structure \citep{Hobler2013,Morrissey2022}. This was determined by the means of molecular dynamics (MD) calculations, which take into account the bonds between atoms. The results have so far only been tentatively confirmed for nepheline \citep[\ch{NaAlSiO4},][]{Martinez2017} where the sputtered secondary \ch{Na+}~ions express a peak in their energy distribution around 2.4~eV, which was attributed to a SBE of Na of 4.8~eV \citep{Morrissey2022}. This exceeds the tabulated value of 1.1~eV by a factor of 4.3. Interestingly, the secondary \ch{K+}~ion results of \cite{Martinez2017} would suggest K SBEs of 4~eV, also exceeding the tabulated value of 0.93~eV by the same factor. \cite{Morrissey2022} also found that within plagioclase---the primary Na bearing mineral on a planetary surface---the surface binding energy is increased to 7.9~eV in the Na end member albite (\ch{NaAlSi3O8}), which would result in a reduction of the Na sputter yield from albite by a factor of 15. The MD results therefore show a positive correlation between SBE and Na coordination number (amount of neighboring atoms).

How the SBE of a damaged surface, or, as outlined by \cite{Hofsass2022}, a non-normal orientation of a mineral unit cell would differ from the ideal conditions chosen in MD simulations is unclear. Furthermore, the energy distributions of secondary ions do not necessarily represent their neutral counterparts, as neutralization of ejected particles is energy-dependent, which can cause a significant offset of the ion distribution towards lower energies \citep{Benninghoven1987,VanderHeide2014}. 
Another example that adds to the uncertainty of the link between neutral and ion energy distributions is from \cite{Betz1987}, who showed that \edit1{ground state Ba sputtered from a continuously oxidized Ba surface coincides with} metastable Ba (originating from the decay of short lived, excited state Ba) and Ba ions \edit1{from a non-oxidized surface}. Ground state Ba \edit1{from a non-oxidized surface} expresses a significantly lower peak energy which can be related to the \edit1{$\Delta H_S$}. The energy distributions of ions, metastable atoms, \edit1{and ground state atoms coincide with each other and exceed $\Delta H_S$}. The larger energy of ions and metastable atoms are interpreted to be caused by matrix dependent ionization processes \citep[e.g.,][]{Dukes2015} \edit1{whereas the increased energy of the sputtered ground state atoms from an oxidized sample are so far not well understood and depend on the procedure including a single initial oxidation or, as in \cite{Betz1987}, a continuous oxidation}. 
What is certain is that the displacement and removal of atoms that would lead to changes in bonds within the sample alters coordination numbers and therefore the binding energy that has to be overcome for their removal. The interatomic potentials between the atoms in the sample would end up far from equilibrium, which is commonly neglected in MD simulations due to computational load \citep{Behrisch2007}. Lastly, \cite{Hobler2013} compared MD and BCA results and concluded, that the enthalpy of sublimation approximation works well in BCA to reproduce experimental data, even when the crystalline structure of the mineral is not taken into account. The reasoning behind this is that in MD simulations, an increase of yield is tied to an increase in defect creation, which ultimately negates the effect of the higher SBEs in the MD simulation. The increased SBEs suggested by MD models are to be taken with caution, but it is established, that an overall increase in energy loss within the sample is necessary to best fit experimental data.

\subsubsection{BB: Bulk binding model}\label{met:BB_model}

The bulk binding (BB) model was recently suggested by \cite{Hofsass2022}. It sets the SBE to zero, whilst setting a BBE for each component which has to be overcome for a component to be freed from their sample and which is lost during each recoil. The authors used the enthalpy of sublimation ($E_s$) for single species samples (i.e., the tabulated values used as SBEs in the surface binding model). For binary compounds, such as oxides and sulfides in minerals, the enthalpy of formation ($\Delta H_f$) has to be overcome before the enthalpy of sublimation of each component, thereby increasing the energy loss in the sample \citep[as suggested earlier by][]{Dullni1984}. 

In SDTrimSP, the implementation of the BB model is similar but slightly different. The sublimation enthalpy of species that form gasses under standard conditions are neglected when determining $E_{bulk}$ (Table~\ref{tab:model_summary}). This is based on the assumption, that, e.g., O from breaking up \ch{SiO2} will already be in its gaseous state and thus will not require to be sublimated, unlike Si. As an example, $E_{bulk}$ (or BBEs) for the elements in the binary compound \ch{SiO2} are, as implemented in SDTrimSP,

\begin{equation}
\begin{aligned}\label{eq:BBE_SiO2}  
    E_{bulk}(\ch{Si}) &= E_s(\ch{Si}) + \frac{\Delta H_f(\ch{SiO2})}{m+n}\\ 
    &= 4.664\,\mathrm{eV} + \frac{9.441\,\mathrm{eV}}{3} = 7.701\,\mathrm{eV}\\
    E_{bulk}(\ch{O}) &= \frac{\Delta H_f(\ch{SiO2})}{m+n}\\
    &= \frac{9.441\,\mathrm{eV}}{3} = 3.147\,\mathrm{eV},
\end{aligned}
\end{equation}
with $m$ and $n$ being the number of components Si and O in the compound (Si$_{m}$O$_{n}$). In SDTrimSP, this model is implemented as the surface-binding-model eight (isbv~=~8), which is only available when using the new density model introduced in Sec.~\ref{sec:compound_density}.

A side-effect of setting the SBE to zero and only using a bulk binding energy (BBE) is a lack of a planar attraction potential and therefore no refraction of sputtered particles towards larger emission angles occurs \citep{Roth1983,Gades1992,Jackson1975,Hofsass2022}. When a surface potential has to be overcome, the extent of the refraction acting on a particle leaving the surface of a sample is proportional to the ratio of the energy of the particle in relation to the potential that has to be overcome \citep{Thompson1968,Sigmund1969}:
\begin{equation}\label{eq:refrac}
    \sin(\theta_1) = \sqrt{\frac{E_0}{E_0-E_\text{sbe}}}\sin(\theta_0),
\end{equation}
with the incident energy $E_0$, the SBE $E_\text{sbe}$, the angle of the atom crossing the surface barrier $\theta_1$, and the initial incident angle of the atom $\theta_0$. Instead, in the BB model, any released particle inside the compound can travel freely through the surface, independent of its energy.

In BCA computations, a cutoff energy ($E_\text{cutoff}$) for each species is set which determines when a recoil is considered to be `at rest' and no longer causes collisions. In the SB model, $E_\text{cutoff}$ is chosen to be 0.1~eV below the lowest, non-zero $E_s$ of all species within the sample. Choosing a lower $E_\text{cutoff}$ would increase computation times due to the impactor travelling deeper into the sample before it is considered at rest. In the context of this work, longer impactor paths are irrelevant because recoils that are below $E_\text{cutoff}$ do not contribute to the sputter yield.
Any recoil from within the sample needs to exceed the SBE to leave the compound with an energy $E_\text{ejecta}$ of
\begin{equation}
E_\text{ejecta} = E_\text{recoil} - \mathrm{SBE}.
\end{equation}
\edit1{This explains why} the $E_\text{cutoff}$ should not be chosen to exceed the SBE of any given component. A recoil of a relatively heavy species that is too slow to overcome the SBE is still capable of causing recoils of lighter species with kinetic energies exceeding their SBE.

For the BB model, however, the BBE is subtracted at each collision, after which recoils can leave the sample without further change of their energy. This energy can therefore be arbitrarily small and has to be limited by the cutoff energy for convergence. With the cutoff, $E_\text{ejecta}$ cannot be inferior to the cutoff energy $E_\text{cutoff}$
\begin{equation}
    E_\text{ejecta} \geq E_\text{cutoff}.
\end{equation}

\begin{figure*}[htbp!]
\gridline{
\fig{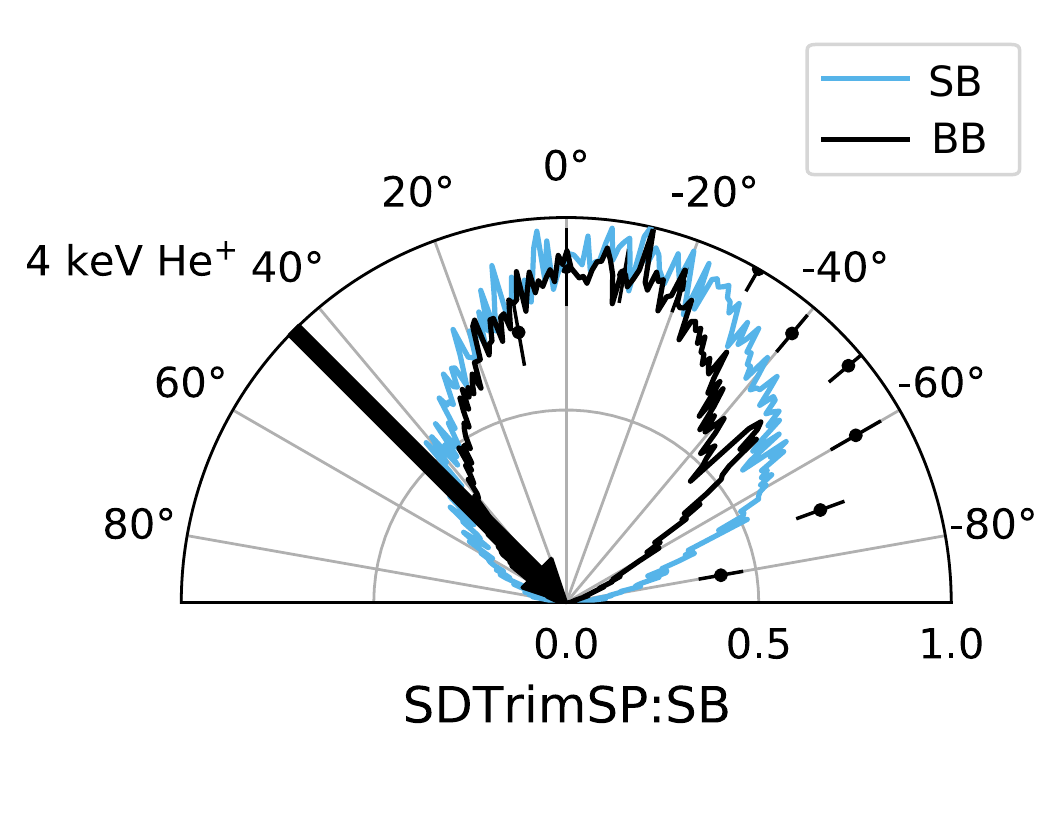}{0.45\textwidth}{}
\fig{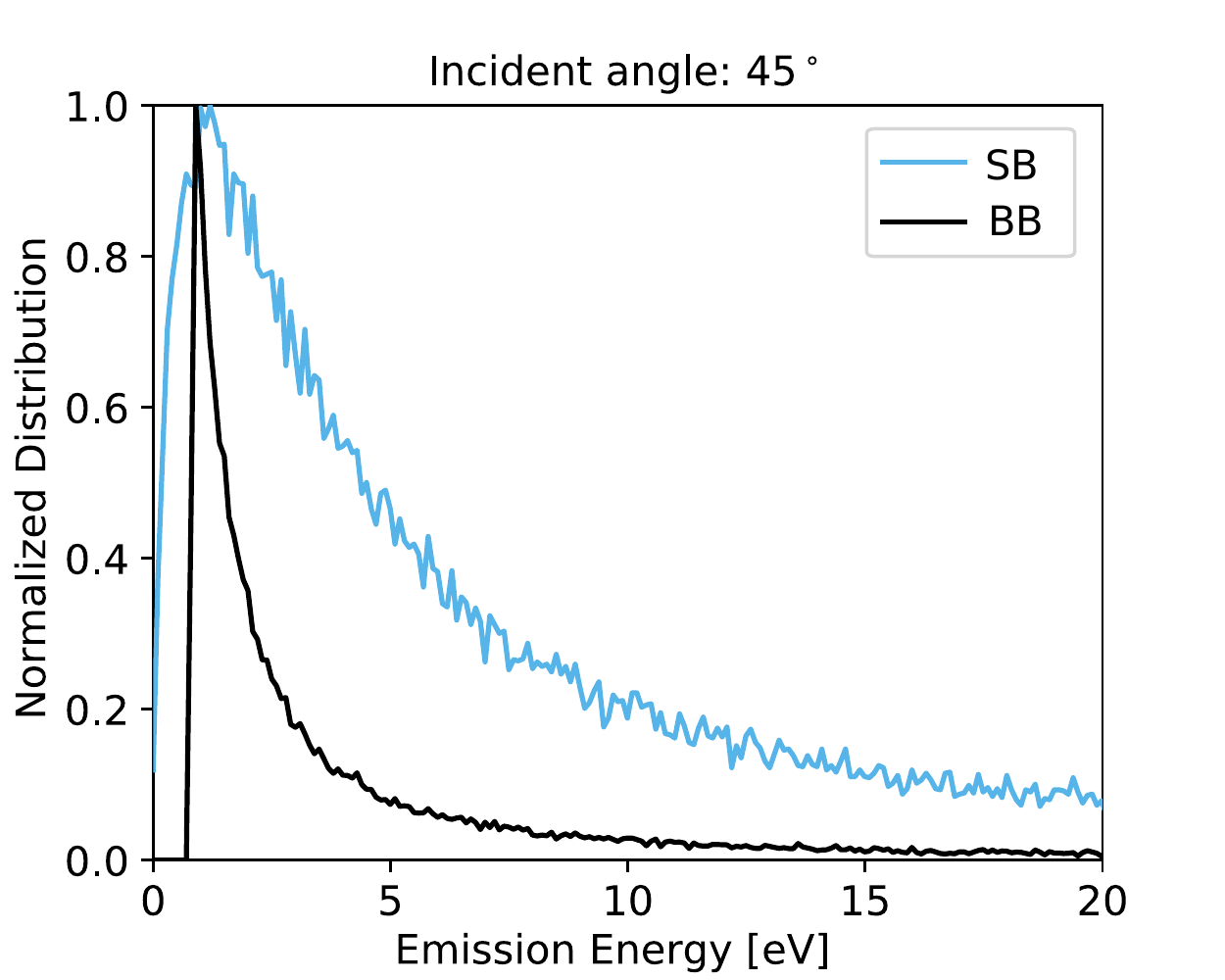}{0.49\textwidth}{}
}
\caption{Model comparison for angular distributions of total sputtered mass yield (left) and energy distribution of sputtered O (right) from irradiated enstatite (\ch{MgSiO3}) for impinging He ions at an incident angle of 45$^\circ$ and energy of 4~keV. The bulk binding model (BB, black line) is based on the pure bulk binding energy (BBE) assumption, where a lack of a surface binding energy (SBE) prevents scattering of the particles towards the surface, resulting in ejecta being preferentially emitted towards the surface normal. The energy distribution of the BB model does not express the characteristic Thompson distribution but instead shows a monotonously decreasing distribution, starting at the element-specific cutoff energy of $\Delta H_s/3$. The surface binding model (SB, light blue line) shown for comparison is calculated with an SBE instead of the BBE. The experimental data are from thin-film irradiation \citep{Biber2022} and normalized to  $y_\text{max}=1$  with an error of one standard deviation.
}
\label{fig:bb_vs_sb}
\end{figure*}

The suggested approach by \cite{Hofsass2022} to obtain the best results to reproduce experimental data is to set a cut-off energy ($E_\text{cutoff}$) in the bulk binding model which lies between 1/2 and 1/8.5 of the atomic $E_s$ (the authors thereby favour a factor of 1/3, which is also the default set for BB models in SDTrimSP). The effect of the absence of a SBE and the use of a BBE and $E_\text{cutoff}$ on the energy distribution of the sputtered particles is evident, as the lower energetic tail of sputtered atoms is cut off at the given $E_\text{cutoff}$, and no Thompson distribution \citep{Thompson1968} is seen (Fig.~\ref{fig:bb_vs_sb}). For the example of \ch{SiO2}, we obtain

\begin{equation}
\begin{aligned}\label{eq:E_cutoff_BBE}
    E_\text{cutoff}(\ch{O}) &= \frac{E_s(\ch{O})}{3} = 0.861\,\mathrm{eV}\\
    E_\text{cutoff}(\ch{Si}) &= \frac{E_s(\ch{Si})}{3} = 1.555\,\mathrm{eV}.
\end{aligned}
\end{equation}

\subsection{HB: New hybrid binding energy model}\label{met:hybrid_model}
The planar potential on the surface is an issue, as its strength needs to exceed atomic enthalpies of sublimation to properly reproduce experimental data. The presence of such a surface potential is however supported by previous energy distribution measurements \citep{Betz1994,Samartsev2006,Martinez2017}. Furthermore, metals covered by a layer of \ch{O2} express energy peak broadening as well as a slight shift to larger energies \citep{Dullni1984,Wucher1986,Wucher1988}. The energy distribution of the BB model is thus only fitting to sputtering of binary metal compounds where monotonously decreasing energy distributions were observed with peak energies close to zero \citep{Szymonski1981}. In oxide-bearing minerals we would thus expect a behavior where the energy distribution is affected proportionally with the amount of available O. Neither the SB nor the BB model is capable to take this into account, which demands a new model.

We introduce a hybrid binding energy model (HB) that uses the element enthalpy of sublimation as SBE and the enthalpy of formation for compounds as BBE. The energies thus represent a surface potential which has to be overcome as well as the bonds within the sample, which have to first be broken up before an atom is mobilized. The model is based purely on tabulated data, just like the bulk binding model of \cite{Hofsass2022} but without the need of a specific $E_\text{cutoff}$ to best reproduce sputter yields and energy distributions. It therefore poses a promising alternative to the previous approaches for obtaining and increased binding energies.

As an example, the SBE and BBE for the binary compound \ch{SiO2} result in
\begin{equation}
\begin{aligned}\label{eq:HB_SiO2} 
    E_{surf}(\ch{Si}) &= E_s(\ch{Si}) = 4.664\,\mathrm{eV}\\
    E_{bulk}(\ch{Si}) &= \frac{\Delta H_f(\ch{SiO2})}{m+n}\\ 
    &=  \frac{9.441\,\mathrm{eV}}{3} = 3.147\,\mathrm{eV}\\
    E_{surf}(\ch{O}) &= \frac{\Delta H_{diss}(\ch{O2})}{2} = 2.582\,\mathrm{eV}\\
    E_{bulk}(\ch{O}) &= \frac{\Delta H_f(\ch{SiO2})}{m+n}\\
    &= \frac{9.441\,\mathrm{eV}}{3} = 3.147\,\mathrm{eV}.
\end{aligned}
\end{equation}

The bulk binding energies which are determined from binary compounds only hold as long as we assume that each element remains bound over the course of irradiation. This is naturally not the case and led in consequence to the implementation of a more sophisticated compound model.

\subsection{New compound model}\label{sec:compound_model}
We propose a simple model for sample compositions which serves two purposes. It allows discrimination between chemically bound atoms and `free' atoms (not chemically bound) and to use data of compounds (i.e., oxides and sulfides) to adequately approximate realistic mass densities of minerals. The simulation names using this compound model to differentiate between bound and un-bound atoms as well as density are labelled by `-C' (HB-C, for the combination of compound and hybrid model; Table~\ref{tab:model_summary}).

\subsubsection{Discriminate between bound and free atoms}\label{sec:compound_discrimination}
Instead of using single atoms, the starting condition considers each atom to be bound to its respective compound---for example, Si and O are bound in \ch{SiO2}. If a recoil occurs with sufficient energy to overcome the bulk binding energy the bound atom is un-bound. The atomic species produced by breaking up compounds have no longer a chemical binding energy (BBE~=~0, Table~\ref{tab:model_summary}). If the remaining energy after the collision is large enough, the target-atom can move trough the sample. The atom then either comes to a halt and attempts to re-form a bond or is ejected. To prevent a major accumulation of atomic species, free atoms react to form the initially set compounds again whenever possible. In the current SDTrimSP implementation, the compound with the highest formation enthalpy is prioritized to re-form given the available O. This has the desired effect that oxygen is unlikely to ever exist as a free atom. In SDTrimSP, the compound hybrid model is implemented as the surface-binding-model four (isbv~=~4). In the non-compound models BB and HB, each component within the sample has a fixed BBE due to the atomic model not being capable of differentiating bound from free components (Table~\ref{tab:model_summary}). They therefore do not behave identical to their compound counterparts (BB-C and HB-C), which causes major differences especially between the HB and HB-C energy and angular distributions (Sec.~\ref{sec:results}).

\subsubsection{Set atomic density with compounds}\label{sec:compound_density}
It was found that the best fitting models to sputter yields for mineral not only require an increase in binding energy \citep[as already hinted at in, e.g.,][]{Dullni1984}, but also an accurate model that reflects realistic material properties which includes the atomic density \citep[e.g.][]{Szabo2020a}. The default way of determining densities in SDTrimSP and TRIM is by using tabulated data of atomic species. In \citep{Szabo2020a}, the authors follow \cite{Moller2001} and calculate a density for wollastonite (\ch{CaSiO3}) based on tabulated atomic densities, which results in 0.0376~atoms~\r{A}$^{-3}$. Increasing the density of oxygen $\rho_\text{O}$ to 0.7~atoms~\r{A}$^{-3}$ (from an initial 0.04~\r{A}$^{-3}$) leads to a bulk density more akin of the wollastonite density of 0.07412~atoms~\r{A}$^{-3}$, corresponding to 2.86~g~cm$^{-3}$. This value for $\rho_\text{O}$ exceeds the typical atomic density by over an order of magnitude. Therefore, in dynamical modeling removal of oxygen causes disproportionate changes to the surface density of the compound compared to removing any other element. To prevent this, we propose calculating mineral densities based on the tabulated atomic densities of compounds, which are simplified building blocks of minerals.

In SDTrimSP, the density of each layer of the sample is calculated based on the density of its components with
\begin{equation}
    \rho=\left(\sum^n_1\frac{X_n}{\rho_n}\right)^{-1},
\end{equation}
where $\rho$ is the density of the sample, $X_n$ the atomic fraction and $\rho_n$ the density of the $n^\text{th}$ component.

The atomic densities and atomic fractions define the bulk density, and therefore the mean free path between two atoms in the sample. The mean free path $\mu$ is formulated in SDTrimSP as
\begin{equation}
    \mu = \rho^{-1/3}.
\end{equation}
In BCA simulations such as SDTrimSP, an ion travelling through the sample will gradually lose its energy through nuclear and electronic interactions, which influence its motion \citep[e.g.,][]{Eckstein1991}. After the impinging ion has travelled the distance $\mu$, a collision occurs \citep{Eckstein1991,Mutzke2019}. High density samples have small $\mu$ and more energy is conserved between two collisions as the effect of electronic stopping is reduced.

Another effect of density is the distance between the atoms and therefore it has an influence on the transferable energy during a collision. This energy is inverse proportional to the distance between the projectile and the center of the particle at rest. The furthest distance at which a collision occurs is the maximal impact parameter, where energy transfer is at its minimum
\begin{equation}
    p_\text{max} = \mu(2\pi)^{-1/2}.
\end{equation}
With smaller $\mu$, the minimum transferable energy becomes larger as the spacing between the atoms, and therefore the mean impact parameter decreases. Higher densities therefore reduce the amount of low-energetic sputtered particles through recoils and lowers the number of recoils as the energy is lost faster.

Mineral densities and calculated mean free paths of relevant rock-forming minerals are shown in Table~\ref{tab:major_minerals}. As an example, for enstatite ($\rho_\text{En} \sim 3.20\;\text{g cm}^{-3}$), the default atomic model would result in 
\begin{equation}
\begin{aligned}
    \rho_\text{En}&=\left(\frac{X_\text{Mg}}{\rho_\text{Mg}}
                    +\frac{X_\text{Si}}{\rho_\text{Si}} +\frac{X_\text{O}}{\rho_\text{O}}\right)^{-1}\\
    \rho_\text{En}&=\left(\frac{0.2}{0.0431}+\frac{0.2}{0.0499} +\frac{0.6}{0.0429}\right)^{-1}\;\text{at A}^{-3}\\ 
                  &=\; 0.0442\;\text{at A}^{-3}\\
                  &= 1.47\;\text{g cm}^{-3},\\
 \end{aligned}
\end{equation}                 
whereas the compound model, using tabulated data for elements results in
\begin{equation}
\begin{aligned}
    \rho_\text{En}&=\left(\frac{X_\text{MgO}}{\rho_\text{MgO}}
                    +\frac{X_\text{SiO2}}{\rho_\text{SiO2}}\right)^{-1}\\
    \rho_\text{En}&=\left(\frac{0.5}{0.1070}
                    +\frac{0.5}{0.0797}\right)^{-1}\;\text{at A}^{-3}\\
                  &=\;0.0913\;\text{at A}^{-3}\\
                  &= 3.05\;\text{g cm}^{-3}.
\end{aligned}
\end{equation}

This example and the results in Table~\ref{tab:major_minerals} demonstrate, how using compound data recreates realistic mineral densities and as a result the mean free path within a sample well. Table~\ref{tab:major_minerals} also shows, that densities can be approximated without any manual adjustments compared to the default atomic model. Together with the hybrid binding energy model, it poses the first step in properly approximating oxides and oxide-derived minerals in Monte Carlo BCA codes such as SDTrimSP.\\

\begin{deluxetable*}{llccccccccc}
	
	\tablecaption{Major rock forming minerals required to represent an unknown planetary surface, consisting of volcanic minerals \label{tab:major_minerals}}

    \tablehead{\colhead{Group} & \colhead{Mineral} & \colhead{Formula} & \multicolumn2c{$\rho_\mathrm{ref}$} & \multicolumn2c{$\rho_\mathrm{compounds}$} &  \colhead{$\Delta\mu_\mathrm{compounds}$} & \multicolumn2c{$\rho_\mathrm{atomic}$} &  \colhead{$\Delta\mu_\mathrm{atomic}$}\\ 
    \colhead{} & \colhead{} & \colhead{} & \colhead{(g/cm$^3$)} & \colhead{(at/\r{A}$^3$)} & \colhead{(g/cm$^3$)} & \colhead{(at/\r{A}$^3$)} & \colhead{(1)} & \colhead{(g/cm$^3$)} & \colhead{(at/\r{A}$^3$)} & \colhead{(1)}} 
    
    \startdata
    Plagioclase & Orthoclase    & \ch{KAlSi3O8}     & 2.56 & 0.0723 & 2.67 & 0.0754 & -1\%  & 1.36 & 0.0384 & 23\%\\
    	        & Albite        & \ch{NaAlSi3O8}    & 2.62 & 0.0786 & 2.70 & 0.0808 & -1\%  & 1.43 & 0.0429 & 22\%\\
    	        & Anorthite     & \ch{CaAl2Si2O8}   & 2.73 & 0.0768 & 2.99 & 0.0840 & -3\%  & 1.53 & 0.0429 & 21\%\\
    	        & Nepheline     & \ch{NaAlSiO4 }    & 2.59 & 0.0747 & 2.84 & 0.0820 & -3\%  & 1.44 & 0.0414 & 22\%\\
    Pyroxene    & Wollastonite  & \ch{CaSiO3}       & 2.93 & 0.0760 & 2.91 & 0.0755 & 0\%   & 1.45 & 0.0375 & 26\%\\
    	        & Diopside      & \ch{CaMgSi2O6}    & 3.40 & 0.0946 & 2.97 & 0.0827 & 5\%   & 1.46 & 0.0405 & 33\%\\
    	        & Enstatite     & \ch{Mg2Si2O6}     & 3.20 & 0.0960 & 3.05 & 0.0913 & 2\%   & 1.47 & 0.0441 & 30\%\\
    	        & Ferrosillite  & \ch{Fe2Si2O6}     & 3.95 & 0.0902 & 3.82 & 0.0872 & 1\%   & 2.15 & 0.0491 & 22\%\\
    Olivine     & Forsterite    & \ch{Mg2SiO4}      & 3.27 & 0.0980 & 3.21 & 0.0960 & 1\%   & 1.46 & 0.0438 & 31\%\\
                & Fayalite      & \ch{Fe2SiO4}      & 4.39 & 0.0908 & 4.64 & 0.0900 & 0\%  & 2.48 & 0.0512 & 21\%\\
    Oxides      & Ilmenite      & \ch{FeTiO3}       & 4.72 & 0.0937 & 4.83 & 0.0959 & -1\%  & 2.54 & 0.0504 & 23\%\\
    	        & Quartz        & \ch{SiO2}         & 2.65 & 0.0797 & 2.65 & 0.0797 & 0\%   & 1.51 & 0.0454 & 21\%\\
    Sulfides    & Troilite      & \ch{FeS}          & 4.61 & 0.0632 & 4.61 & 0.0632 & 0\%   & 3.89 & 0.0533 & 6\%\\
            	& Niningerite   & \ch{MgS}          & 2.68 & 0.0573 & 2.68 & 0.0573 & 0\%   & 1.91 & 0.0408 & 12\%\\
    	        & MnS           & \ch{MnS}          & 3.99 & 0.0552 & 3.99 & 0.0552 & 0\%   & 3.80 & 0.0526 & 2\%\\
    	        & CrS           & \ch{CrS}          & 4.89 & 0.0701 & 4.89 & 0.0701 & 0\%   & 3.70 & 0.0530 & 10\%\\
    	        & TiS           & \ch{TiS}          & 3.85 & 0.0580 & 3.85 & 0.0580 & 0\%   & 3.07 & 0.0462 & 8\%\\
    	        & CaS           & \ch{CaS}          & 2.59 & 0.0432 & 2.59 & 0.0432 & 0\%   & 1.74 & 0.0290 & 14\%\\
    Accessories & Spinel        & \ch{MgAl2O4}      & 3.64 & 0.1078 & 3.77 & 0.1115 & -1\%  & 1.58 & 0.0468 & 32\%\\
    	        & Chromite      & \ch{FeCr2O4}      & 4.79 & 0.0902 & 5.29 & 0.0996 & -3\%  & 2.88 & 0.0543 & 18\%\\
    \enddata
    \tablecomments{Difference in mean free path lengths ($\mu = \rho^{-1/3}$) are calculated as $\Delta\mu = \mu/\mu_\mathrm{ref}-1$; The density short forms are: $\rho_\mathrm{ref}$ -- mass densities and atomic densities calculated based on typical mineral densities found on \hyperlink{http://webmineral.com}{webmineral} \citep[see also, e.g.,][]{Deer1992}; $\rho_\mathrm{compounds}$ -- densities calculated based on tabulated oxide and sulfide data from pure compound properties; $\rho_\mathrm{atomic}$ -- densities calculated based on atomic data included in tables of SDTrimSP which are based on mono-atomic solids.}

\end{deluxetable*}  

\begin{deluxetable*}{@{\extracolsep{4pt}}lccccccc@{}}
	\renewcommand{\arraystretch}{1.3}
	\tablecaption{The different energy and density models and their parameters \label{tab:model_summary}}

    \tablehead{\colhead{} & \multicolumn{5}{c}{SDTrimSP model presets} & \multicolumn{2}{c}{manually set models} \\
    \cline{2-6} \cline{7-8}
    \colhead{} & \colhead{SB} & \colhead{SB-C} & \colhead{BB}  & \colhead{BB-C} & \colhead{HB-C}& \colhead{BB$_0^*$}& \colhead{HB$^*$}}
    \startdata
    $SBE$                   & $\Delta H_\text{sub}$ & $\Delta H_\text{sub}$ &0&0& $\Delta H_\text{sub}$&0& $\Delta H_\text{sub}$ \\
    $BBE_\text{f}^\dagger$ &0&0& $\Delta  H_\text{sub}$ &0&0& $\Delta H_\text{sub}$ + CBE & CBE \\
    $BBE_\text{b}^\dagger$ & -                     &0& -              & $\Delta H_\text{sub}$ + CBE & CBE& -                            & - \\
    $\rho_\text{f}$                 & atomic                & atomic                & atomic         & atomic & atomic& atomic                       & atomic \\
    $\rho_\text{b}$                 & -                     & compound              & -              & compound & compound& -                            & -  \\
    $E_\text{cutoff}$               & $<\,\Delta H_\text{sub}$  & $<\,\Delta H_\text{sub}$  & $\Delta H_\text{sub}/3$ & $\Delta H_\text{sub}/3$ & $<\,\Delta H_\text{sub}$& $\Delta H_\text{sub}/3$ & $<\,\Delta H_\text{sub}$ \\
    isbv                            &1&1&8&8&4&1&1  \\
    \enddata
    \tablecomments{Short forms:
    SBE -- surface binding energy; 
    BBE -- bulk binding energy;
    f -- `free', un-bound atom; 
    b -- compound-bound atom;
    CBE -- Chemical Binding Energy: $\Delta H_f/(m+n)$ whereas m and n are the number of cations and anions in a compound;
    $E_\text{cutoff}$ -- Cutoff energy;
    $\Delta H_\text{sub}$ -- enthalpy of sublimation;
    $\Delta H_f$ -- enthalpy of formation of binary compound;
    isbv -- model number in SDTrimSP input files.\\
    $^*$Each component is considered un-bound in regards to its density and bound regarding to the BBE (CBE assigned). The BB$_0$ model is the original \cite{Hofsass2022} model. The HB model is only used to demonstrate the effect of density independent of the hybrid binding energy model. \\
    $^{\dagger}$For O, $\Delta H_\text{sub}$ is neglected and only CBE is used as a BBE, if any.\\
}
\end{deluxetable*}

\subsection{Fitting the simulated data}\label{sec:fitting}

The modeled sputter yield by element and mass is fitted using an Eckstein fit  based on the \cite{Yamamura1983} formula \citep{Eckstein2003}:
\begin{equation}
\label{eq:eckstein}
\begin{aligned}
      Y(\upalpha) = 
      & Y(0)\left\{\mathrm{cos}\left[\left(\frac{\upalpha}{\upalpha_0}\frac{\pi}{2}\right)^c\right]\right\}^{-f}\times\\
      &\mathrm{exp}\left\{b\left(1-1\bigg/\mathrm{cos}\left[\left(\frac{\upalpha}{\upalpha_0}\frac{\pi}{2}\right)^c\right]\right)\right\},
\end{aligned}
\end{equation}
with the fitting parameters $b$, $c$, and $f$ and the angle of incidence $\upalpha$. The value for $\upalpha_0$ is chosen as $\pi/2$ instead of being calculated by
\begin{equation}
    \label{eq:yield_adependency}
    \upalpha_0=\pi-\mathrm{arccos}\sqrt{\frac{1}{1+E_0/E_{sp}}}\geq\frac{\pi}{2},
\end{equation} 
because the projectile binding energy $E_{sp}$ would be required or assumed and for the typical solar wind energies $E_0$ in keV range with $E_{sp}$ in the eV range this would cause only minor deviations from $\upalpha_0=\pi/2$.

For the angular distribution of sputtered particles, the data are fitted using an adapted cosine fit function after \citep{Hofsass2022} to take the non-symmetrical nature of sputtered particles into account. The system of equation is as follows:
\begin{equation}
    \label{eq:cosine}
    f(\phi) 
     \left\{
    \begin{aligned}
        & A\,\cos^m\left(\frac{\pi}{2}\left(\frac{\pi+2\phi}{\pi+2\phi_\text{tilt}}-1\right)\right)\;&\phi\leq\phi_\text{tilt} \\
        & A\,\cos^n\left(\frac{\pi}{2}\left(1-\frac{\pi-2\phi}{\pi-2\phi_\text{tilt}}\right)\right)\;&\phi\geq\phi_\text{tilt},
    \end{aligned}
    \right.
\end{equation} 
with the scaling factor $A$, the tilt angle $\phi_\text{tilt}$, the exponents $m$ and $n$, and the angle $\phi$.

The energy distribution data are fitted using a Thompson distribution \citep{Thompson1968},
\begin{equation}
   \label{eq:thompson}
   f(E) =  S\frac{E}{(E + E_0)^3}
   ,
\end{equation} 
with a scaling factor $S$, the energy removed from the sputtered atom before it escapes the surface $E_0$ (approximately SBE, when considering a pure SB model) and the energy of the sputtered atom $E$. The energy peak is located at $E\approx E_0$/2. 


\section{Results} \label{sec:results}
The validity of any new suggested model can ultimately only be verified through experimental data focusing on speciation of the sputtered material as well as its angular and energy distribution. For now, we can only compare experimental sputter yield data in mass per impinging ion (amu/ion) and their angular distribution with model outputs. 
The composition of the modeled yield is stoichiometric. Lighter species are initially sputtered in an over-stoichiometric way. With fluence and decreasing abundance of light species, the sputter yield composition approaches the initial sample stoichiometry, which evidently will not correspond to the sample surface composition in equilibrium. We know that the laboratory data correspond to fluences where this irradiation-equilibrium is reached. For the scope of this work, we assume that the laboratory yield composition is indeed stoichiometric. 

\subsection{HB-C model and experimental data}
\begin{figure}[htbp!]
\gridline{
\fig{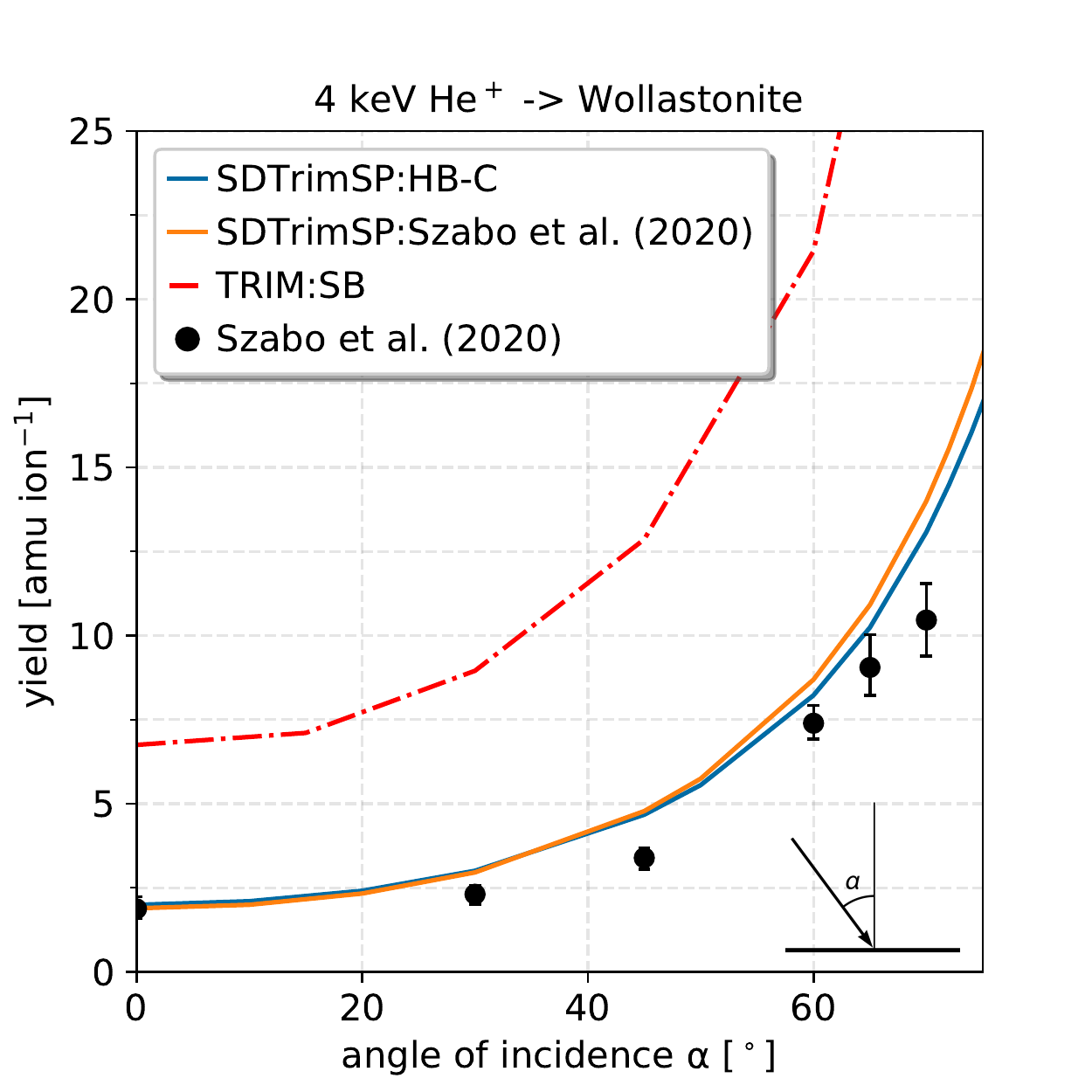}{0.49\textwidth}{}
}
\caption{The agreement of the initial approach used to fit the experimental data \citep{Szabo2020a} with the HB-C model is shown, including TRIM model results \cite{Biersack1984}. The abbreviations are: HB -- Surface binding energy (SBE) based on heat of sublimation and bulk binding energy on enthalpy of formation; C -- densities calculated based on compound densities and differentiation between unbound and bound species. \cite{Szabo2020a} used an averaged SBE of all components after increasing the O\textsubscript{SBE} to 6.5~eV. To reach the proper wollastonite density, they increased the O atomic density accordingly. \label{fig:SzaboVSHybrid}}
\end{figure}

\begin{figure*}[!ht]
    \centering
    \gridline{
    \fig{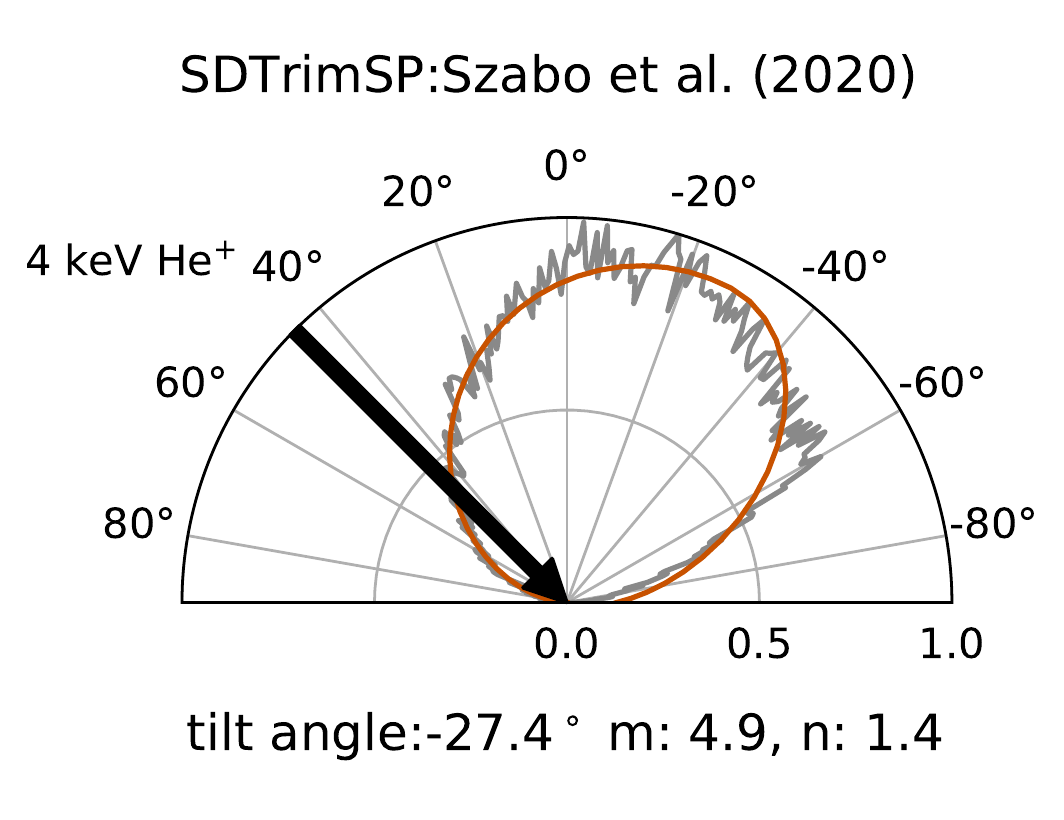}{0.43\textwidth}{}
    \fig{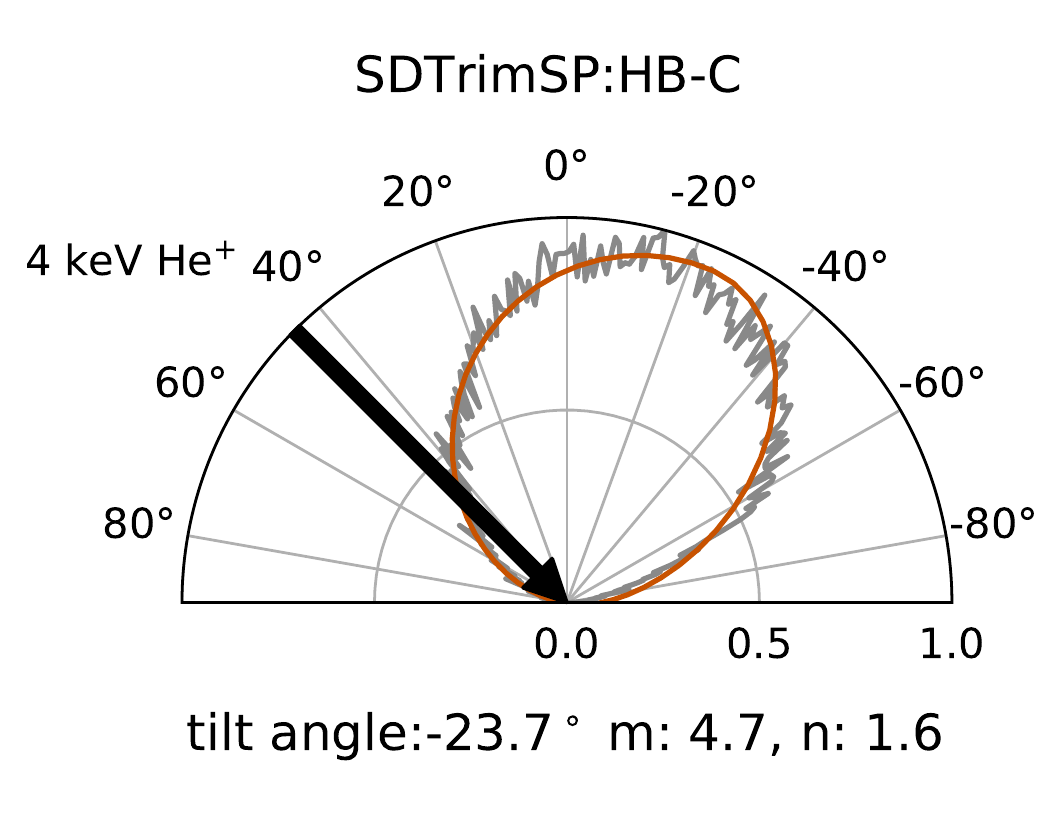}{0.43\textwidth}{}
    }
    \gridline{
    \fig{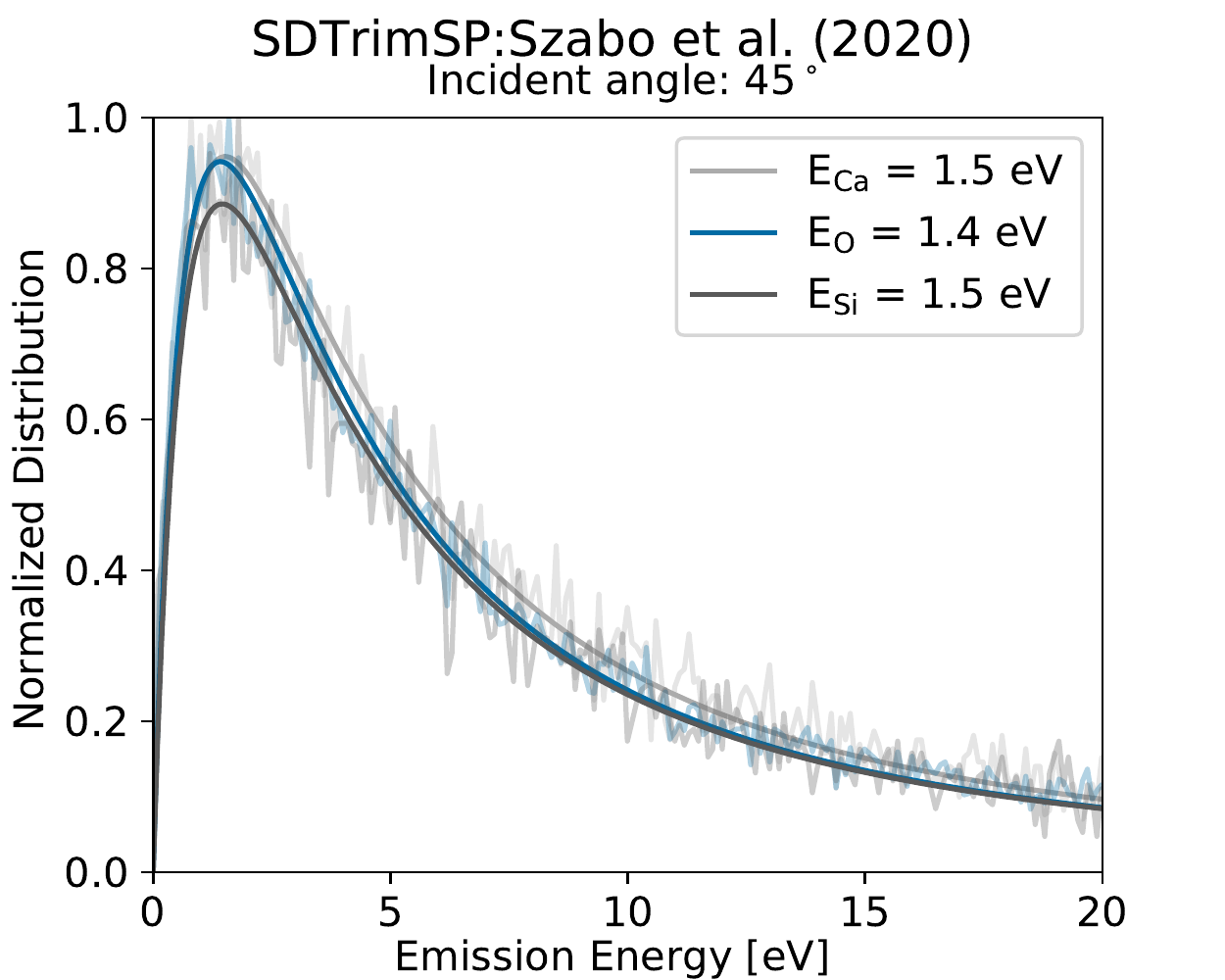}{0.49\textwidth}{}
    \fig{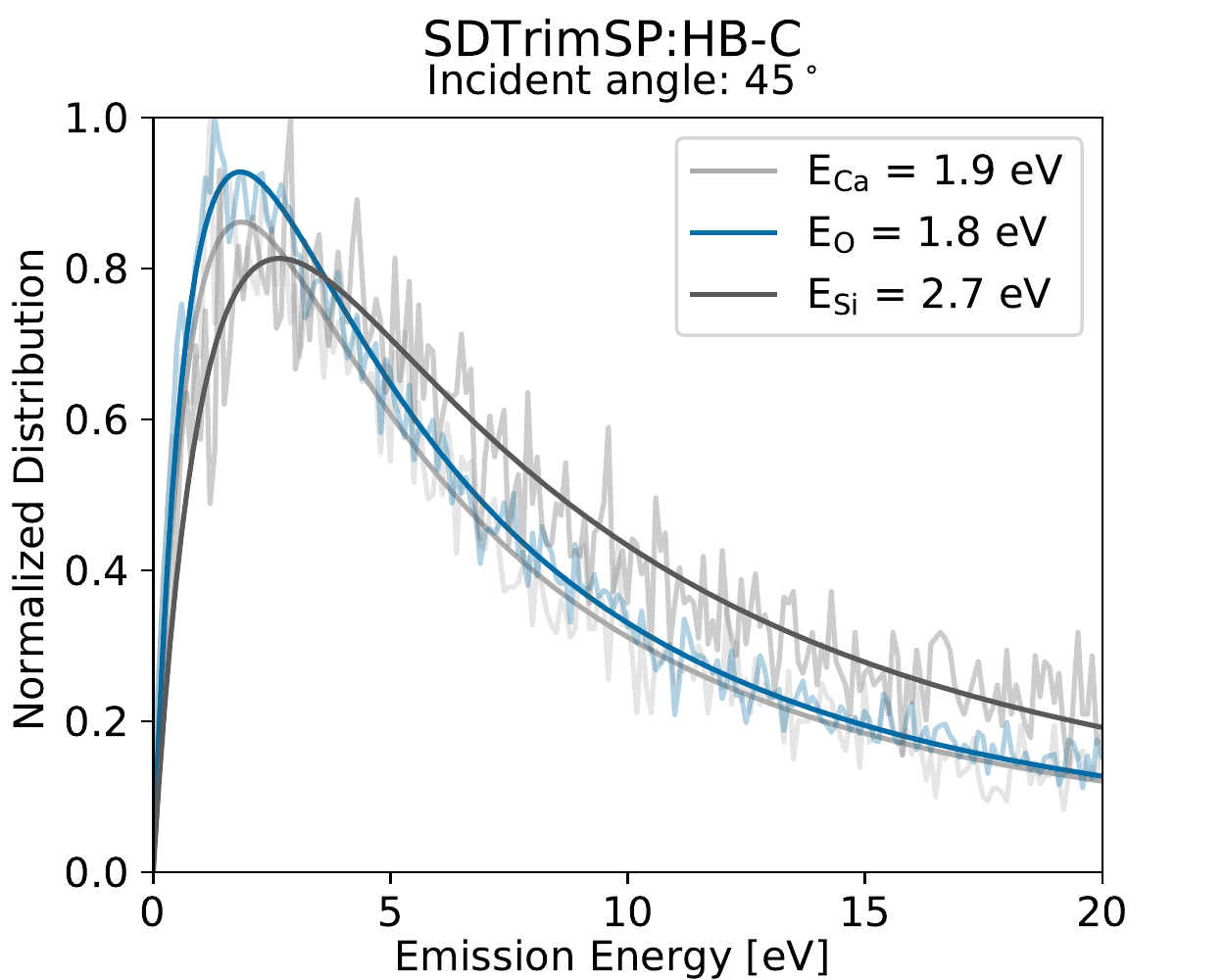}{0.49\textwidth}{}
    }
    \caption{Modeled angular distribution of total sputter yield (data in grey, fit in orange) and energy distributions of sputter ejecta The energy in the legend corresponds to the peak energy of the Thompson fit function. from wollastonite irradiated by 4~keV~\ch{He+}. \cite{Szabo2020a} increased the O surface binding energy (SBE) to 6.5~eV, averaged SBEs for all elements, and increased O density to reach initial wollastonite density. The large surface binding energy causes a high degree of surface scattering of the ejected particles whereas the averaging of the binding energies leads to an identical energy distribution for all species. The HB-C model uses both SBE and bulk binding energy to achieve an increase in binding energy whilst reliably reproducing mineral densities based on oxide compound data and differentiating between compound-bound and un-bound atoms.
    \label{fig:SzaboVSHybrid_dist}
    }
\end{figure*}

We first present the capabilities of the newly implemented hybrid binding energy model which includes the compound model (HB-C). The results of the \cite{Szabo2020a} approach and the HB-C model are thereby largely identical when it comes to mass yields and recreate the experimental data reasonably well (Fig.~\ref{fig:SzaboVSHybrid}). The largest discrepancies lie in both the angular and energy distributions. A high SBE increases the refraction which occurs on the surface, and therefore increases the spread of the angular distribution. We show this behavior in Fig.~\ref{fig:SzaboVSHybrid_dist} where the \cite{Szabo2020a} approach---with the highest SBEs of all model results shown in this work---leads to the largest tilt angle (27$^\circ$ at an angle of incidence of 45$^\circ$) with the broadest angular distribution of all models (exponents $m=4.9$ and $n=1.4$ for \ch{He+} on wollastonite). The homogeneous, atom-insensitive energy distribution of the \cite{Szabo2020a} approach is the consequence of using an identical SBE for each species (Fig.~\ref{fig:SzaboVSHybrid_dist}).

\subsection{All model comparison}
In Fig.~\ref{fig:WoEnVSexp} we compare the HB-C model with other models in relation to the experimental sputter yield data of wollastonite and enstatite. It is apparent that we find the experimental data lying between the HB-C and the HB model. The latter thereby does not differentiate between bound and unbound species in the sample. Most relevant is that the experimental data are recreated using the HB-C model at normal incidence and close to normal incidence ($<45^\circ$).

\subsubsection{Angular distributions}
We compare to experimental angular distributions of \cite{Biber2022} with modeled data of enstatite in Fig.~\ref{fig:adist_En}. The largest agreement with experimental data is with the HB model, which expresses the strongest degree of forward sputtering (largest tilt angle) due to the high binding energy of each species in the sample. The cases with lower or no BBE---this includes the unbound species within the HB-C model---clearly show a drastically reduced degree of forward sputtering compared to the HB model. Angular distribution data of TRIM is not shown, as it expresses distributions even more narrow than the BB model \citep[Fig.~\ref{fig:bb_vs_sb}][]{Hofsass2022}.

\subsubsection{Energy distributions}
Although no experimental data exists for the irradiated enstatite, we present the modeled energy distributions of the sputter ejecta in Fig.~\ref{fig:edist_En}. The SB and SB-C model show a nearly identical energy distribution, whilst the HB and HB-C models express a smaller amount of low energy particles and thus broader peaks. The more prominent, high energy tail of sputtered particles in the HB model is due to the species experiencing large BBEs at any degree of applied fluence. In comparison, the compound model (HB-C) can build up free Mg which are consecutively sputtered without having to overcome a BBE. This in return increases the number of low energy Mg in the energy distribution, which lies closer to the SB-C model. This is manifested in the Mg energy distribution peaking at 0.9~eV in the HB-C model compared to the 0.6~eV in the SB models and the 1.8~eV in the HB model.

\begin{figure*}[!ptbh]
    \gridline{
    \fig{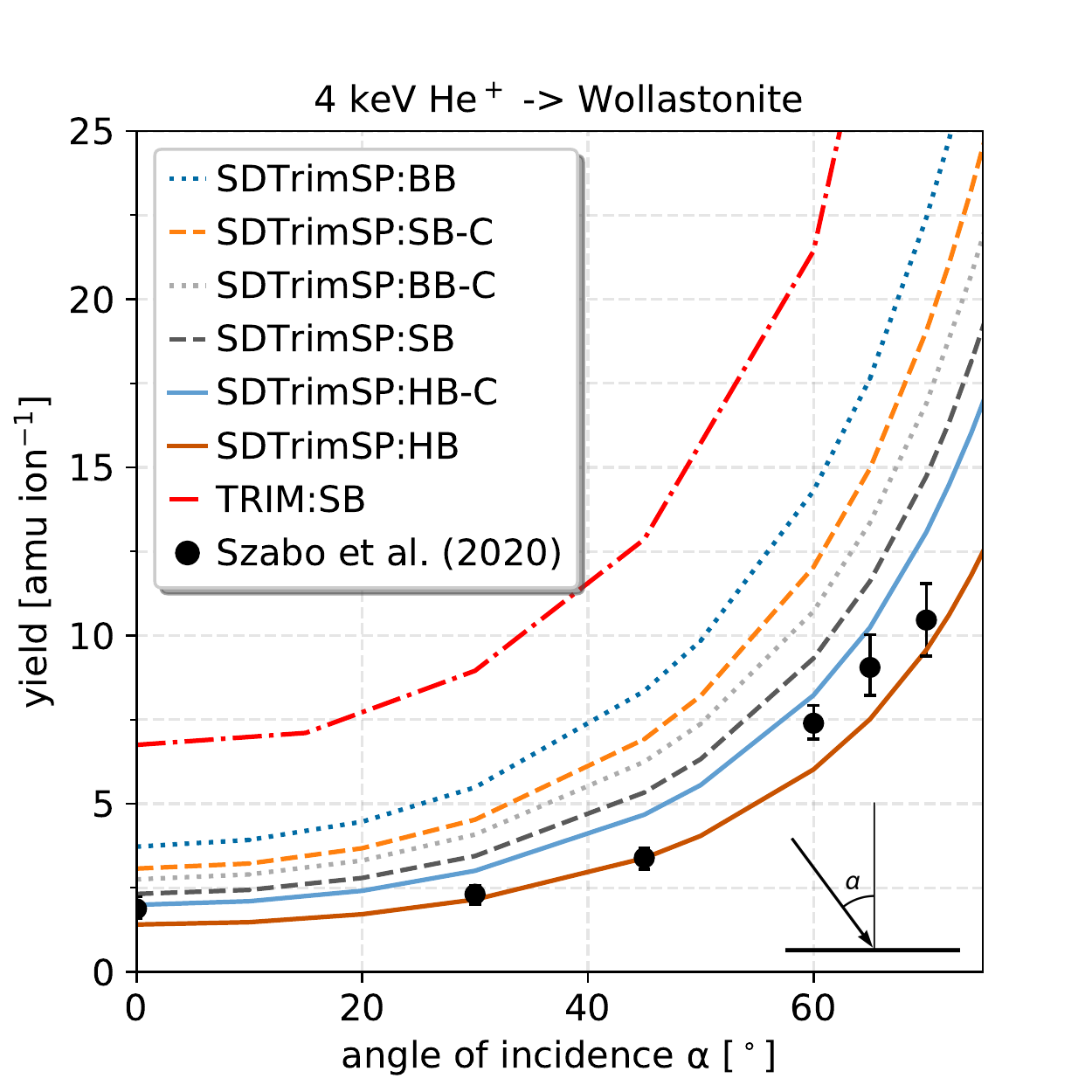}{0.50\textwidth}{}
    \fig{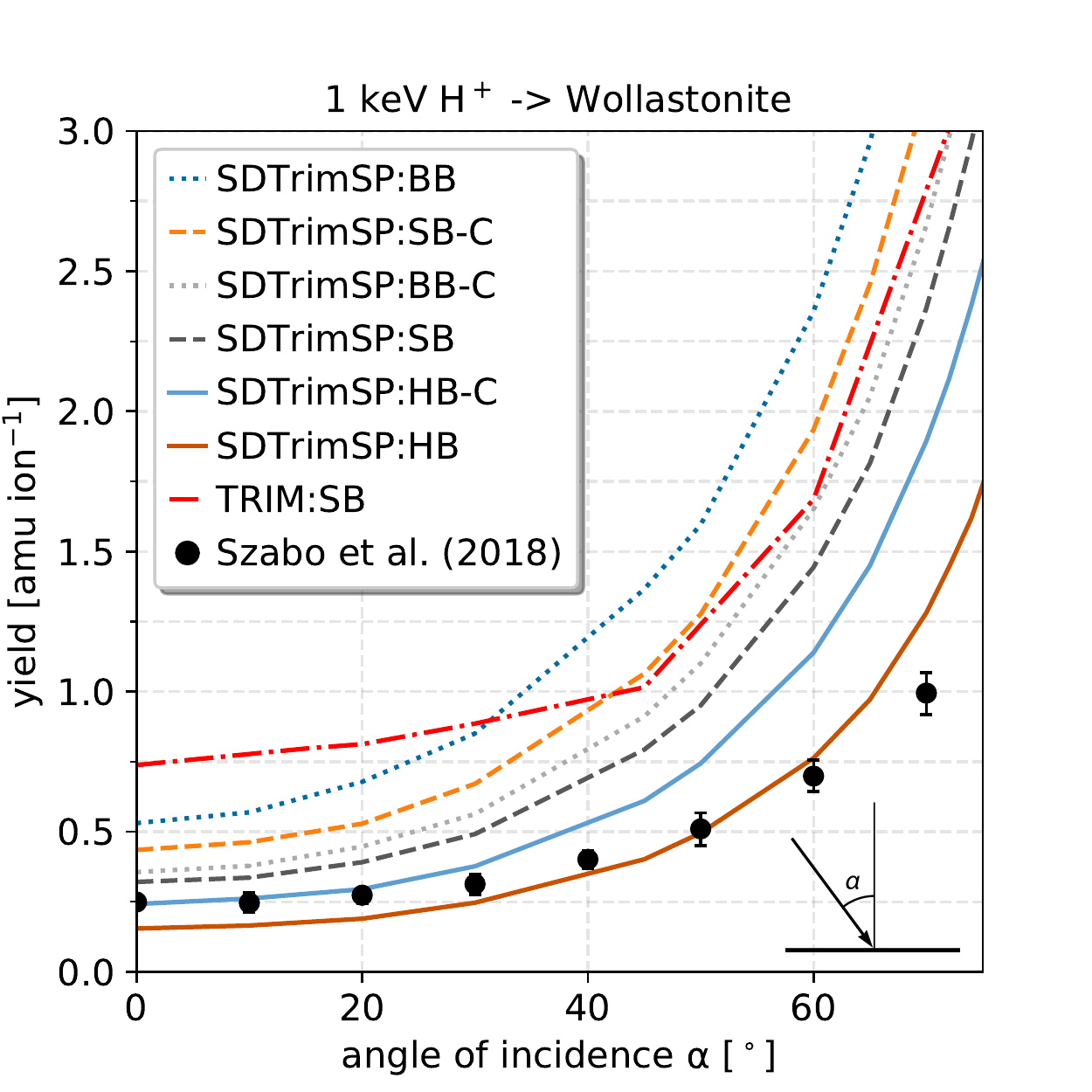}{0.50\textwidth}{}
    }
    \gridline{
    \fig{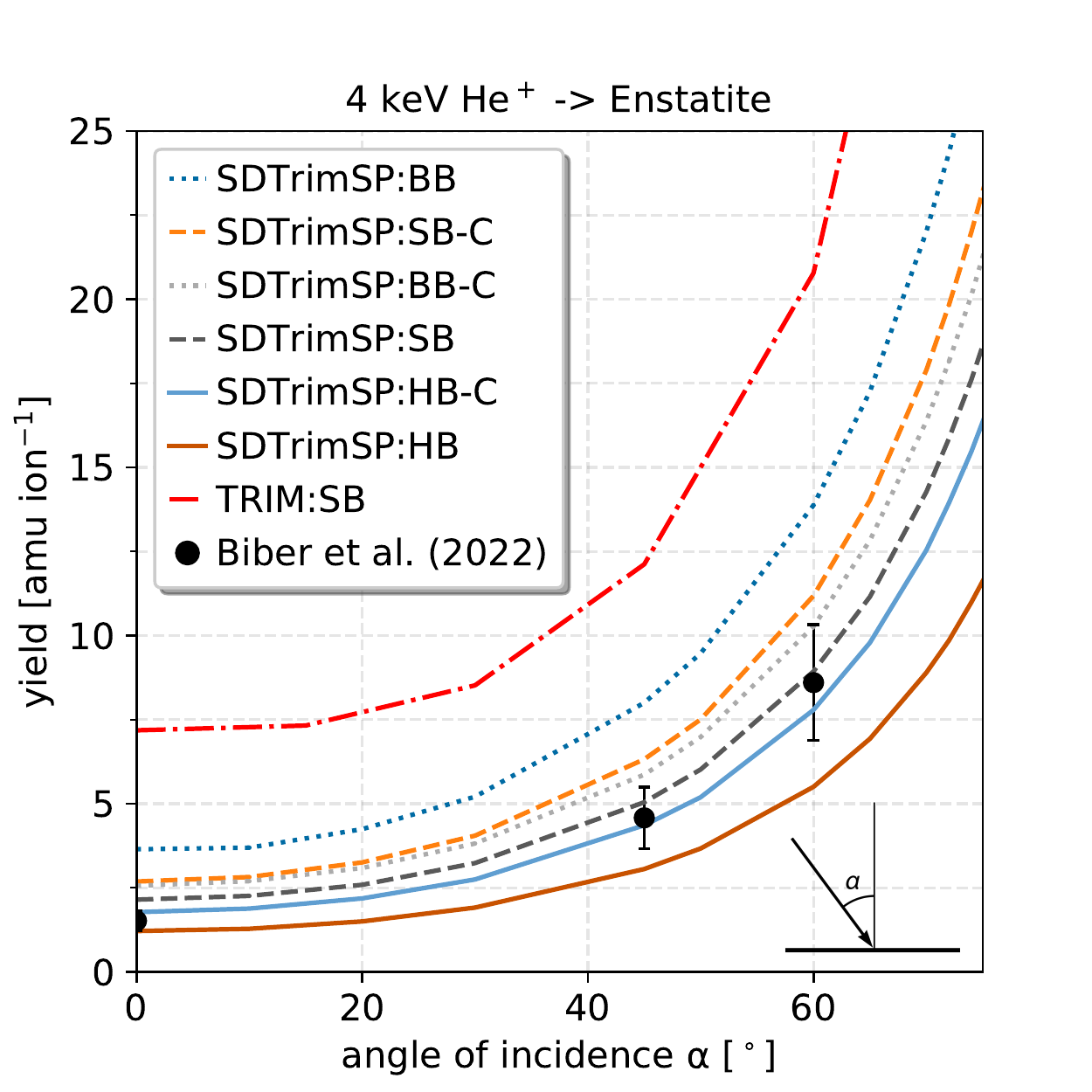}{0.50\textwidth}{}
    }
    \caption{SDTrimSP model results compared to TRIM model results \citep[red dash-dotted line][]{Biersack1984} and experimental data by \cite{Szabo2018} (\ch{H+} on wollastonite), \cite{Szabo2020a} (\ch{He^+} on wollastonite) and \cite{Biber2022} (\ch{He^+} on enstatite). Near ideal mineral densities are obtained in models taking compounds (-C) into account whereas the atomic cases represent lower densities, about a factor two below compound derived densities. Abbreviations and line styles: SB~--~dashed lines, tabulated enthalpy of sublimation as element surface binding energies; BB~--~dotted lines, tabulated enthalpy of sublimation as element bulk binding energies; HB~--~solid lines, tabulated enthalpy of formation as bulk binding energy and enthalpy of sublimation as surface binding energies; C~--~densities calculated based on compound densities and differentiation between compound-bound and un-bound atoms.
    \label{fig:WoEnVSexp}}
\end{figure*}

\begin{figure*}[!ptbh]
    \centering
    \gridline{
    \fig{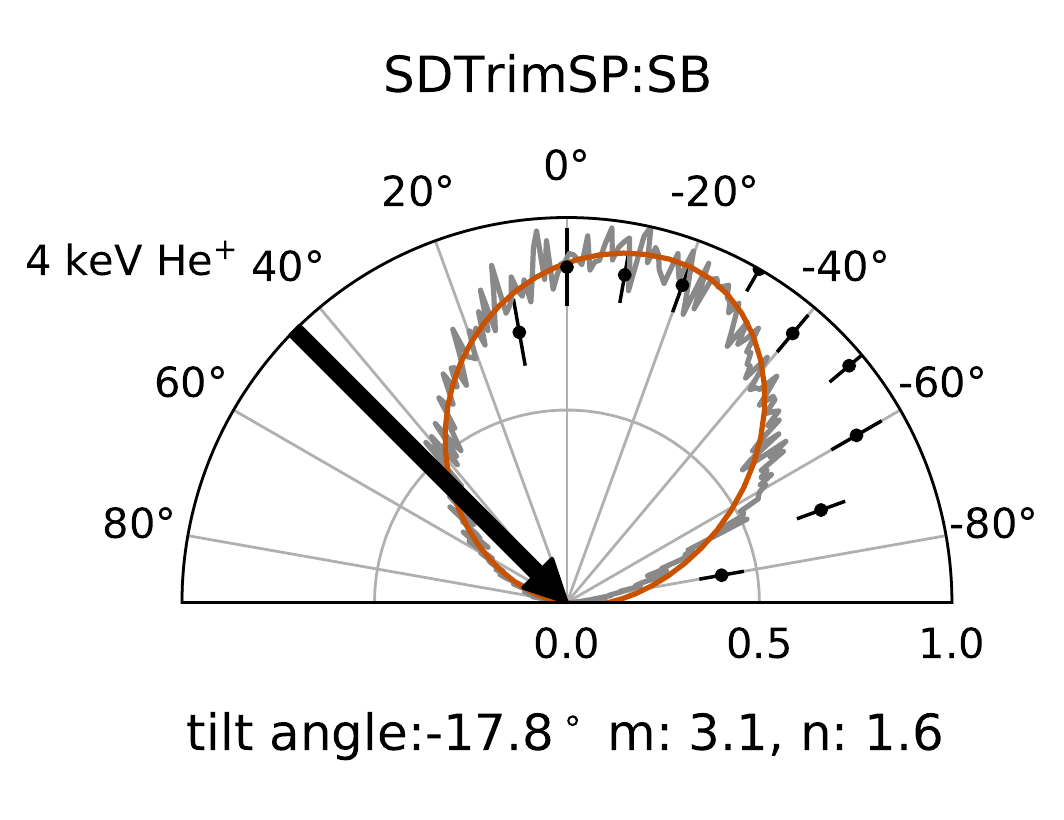}{0.43\textwidth}{}
    \fig{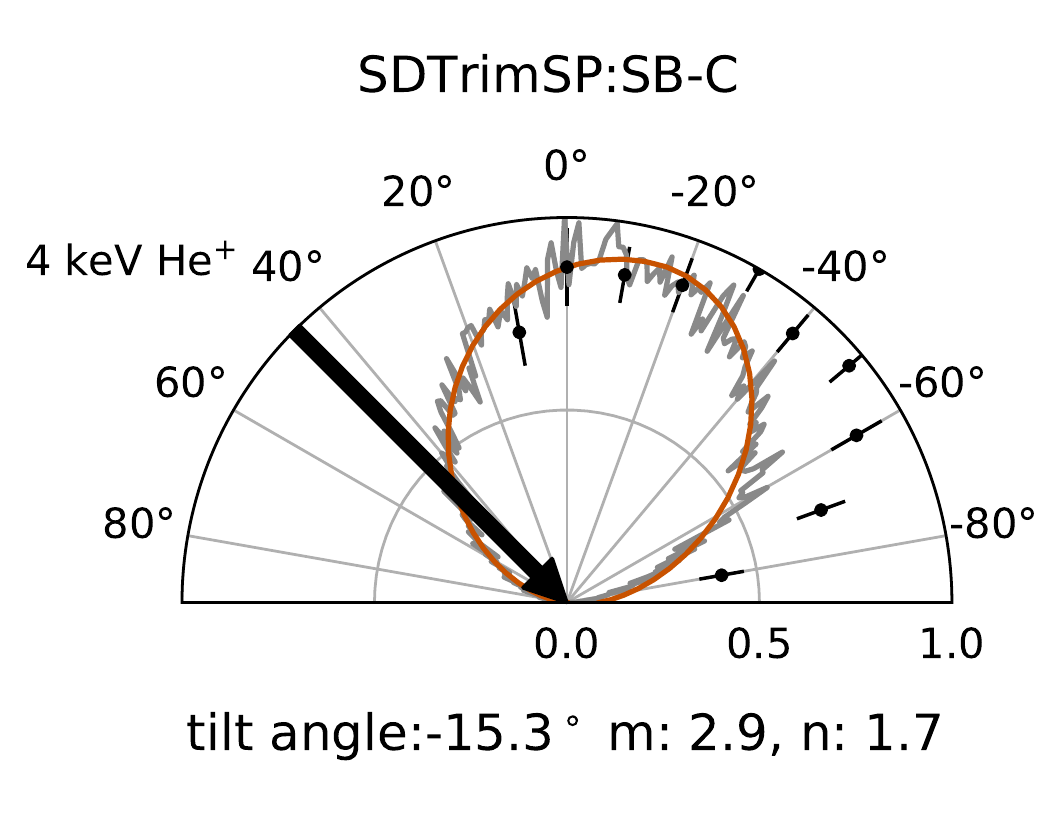}{0.43\textwidth}{}
    }  
    \gridline{
    \fig{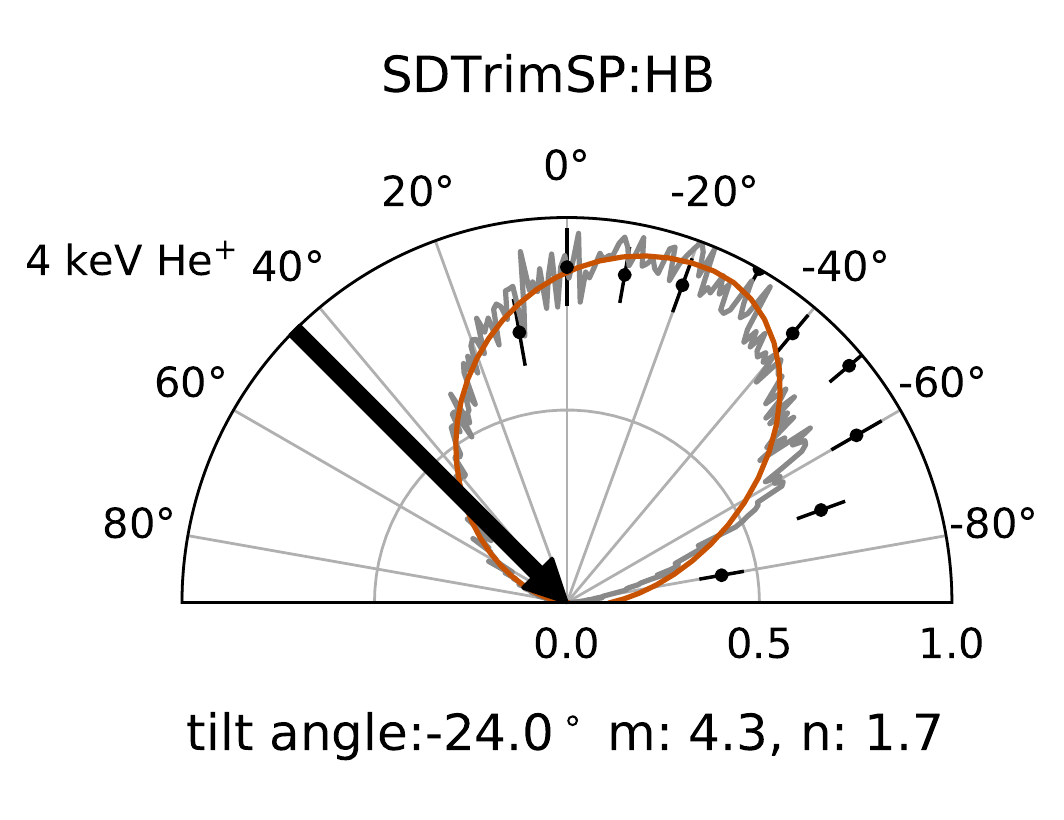}{0.43\textwidth}{}
    \fig{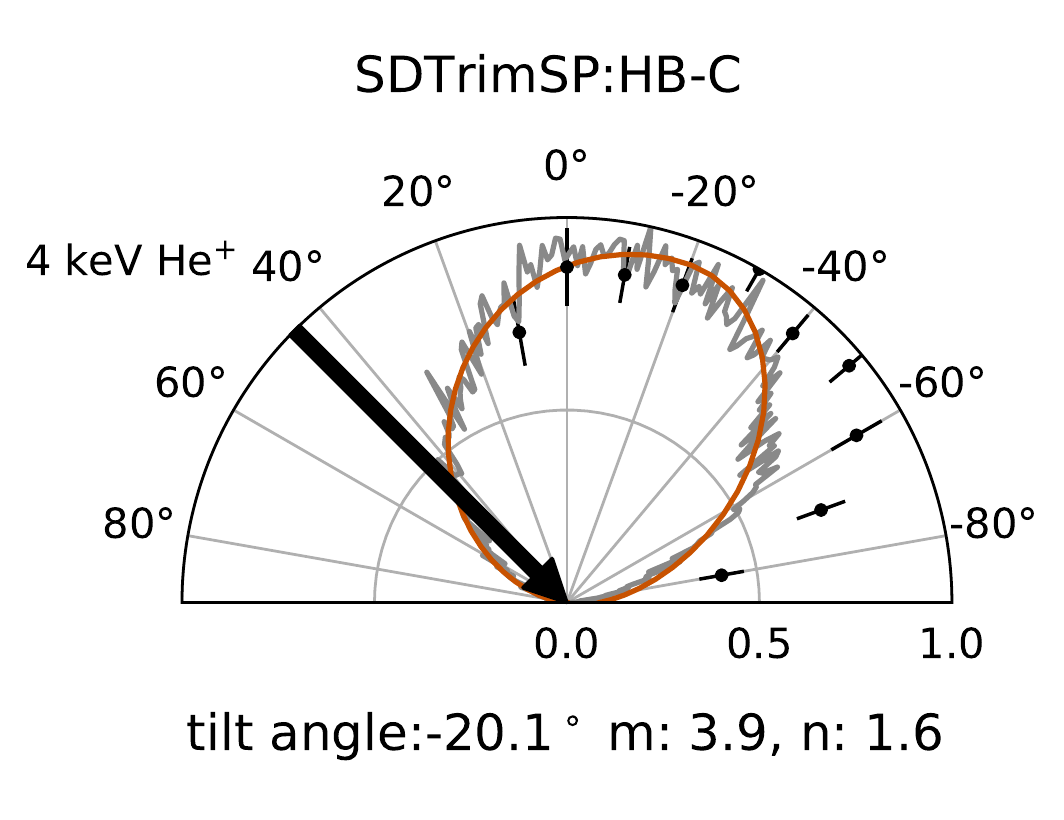}{0.43\textwidth}{}
    }
    \caption{Polar angular distributions of total sputter yields from enstatite irradiated with 4~keV~\ch{He+} at an angle of 45$^\circ$ based on different model assumptions. The larger density prescribed by the compound model leads to a slightly more narrow angular distribution---seen in the smaller $m$ fit exponents of 2.9 and 3.9 of the cosine fit---when compared to the atomic model $m$ exponents of 3.1 and 4.3 respectively. If elements become un-bound with irradiation (HB-C model), the effect of a bulk binding energy (BBE) on the tilt angle is small compared to the SB model ($+2.3^\circ$). If elements remain bound and experience a constant BBE and surface binding energy (HB model), forward sputtering is more prominent (SB model tilt $+6.2^\circ$).  Abbreviations: SB--tabulated enthalpy of sublimation as element surface binding energies;  HB -- tabulated enthalpy of formation as bulk binding energy and enthalpy of sublimation as surface binding energies; C -- densities calculated based on compound densities and differentiation between compound-bound and un-bound atoms. Experimental data from thin-film irradiation \citep{Biber2022} normalized to  $y_\text{max}=1$  with an error of one standard deviation. 
    \label{fig:adist_En} 
    }
\end{figure*}

\begin{figure*}[!tbhp]
    \gridline{
    \fig{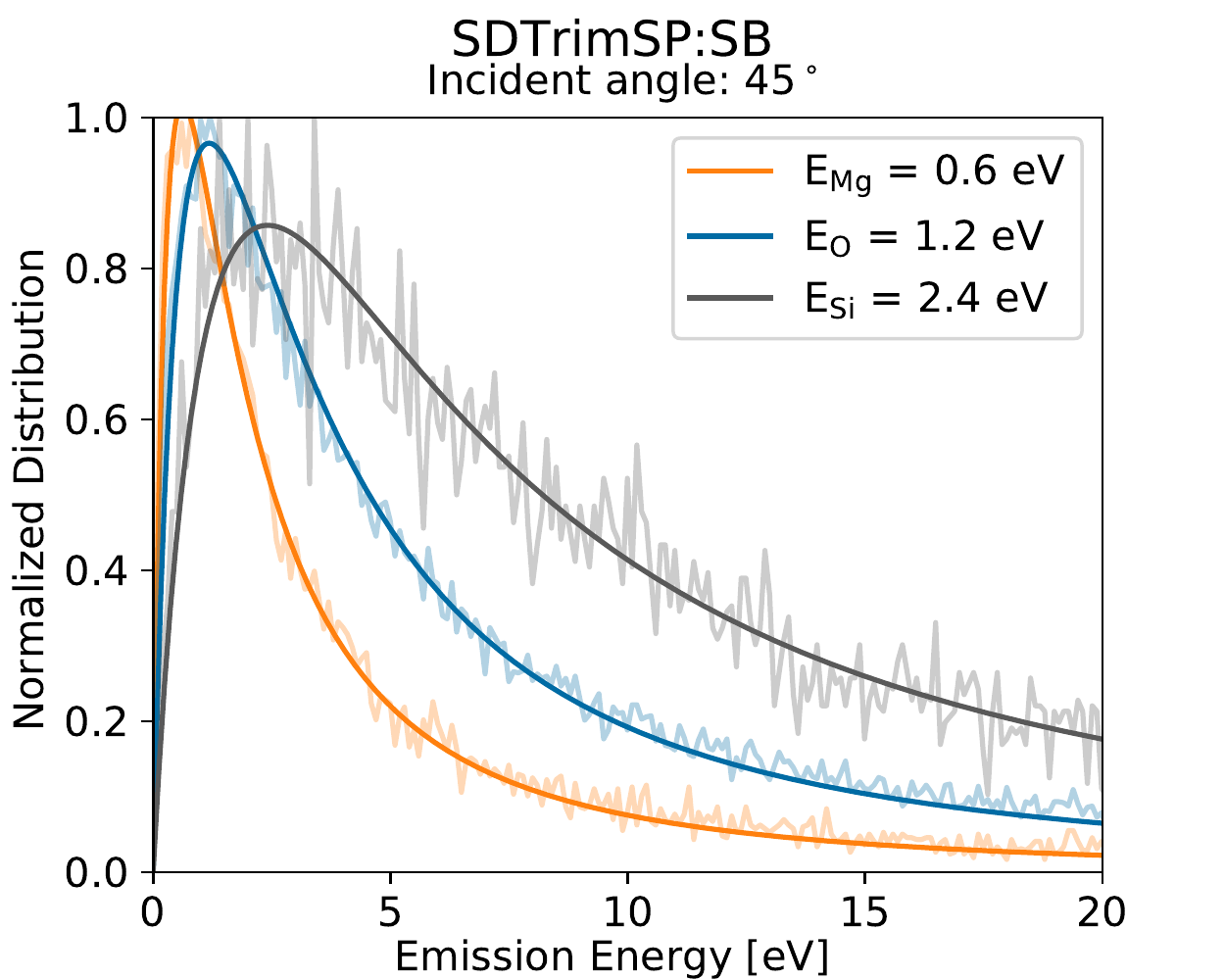}{0.49\textwidth}{}
    \fig{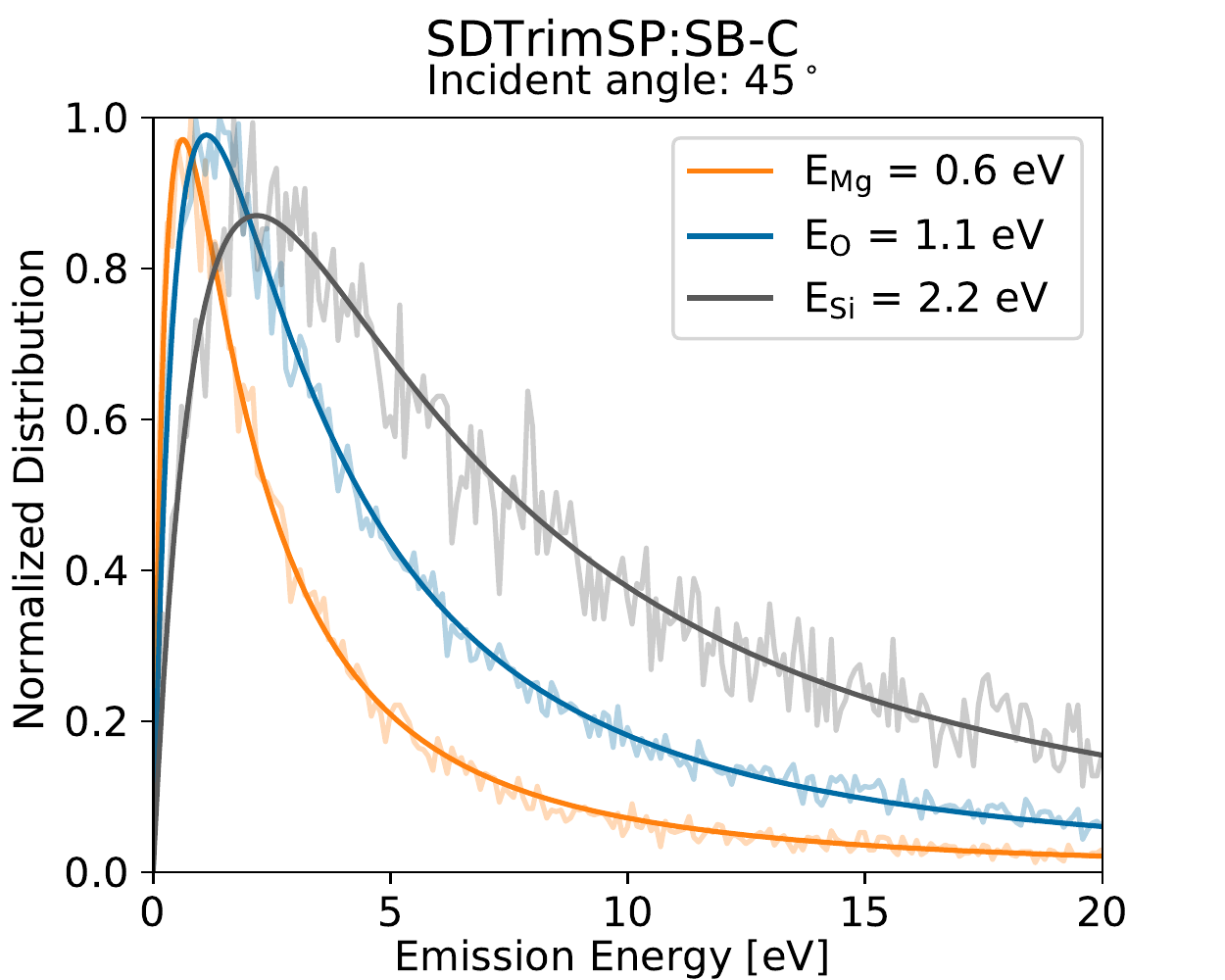}{0.49\textwidth}{}
    }    
    \gridline{
    \fig{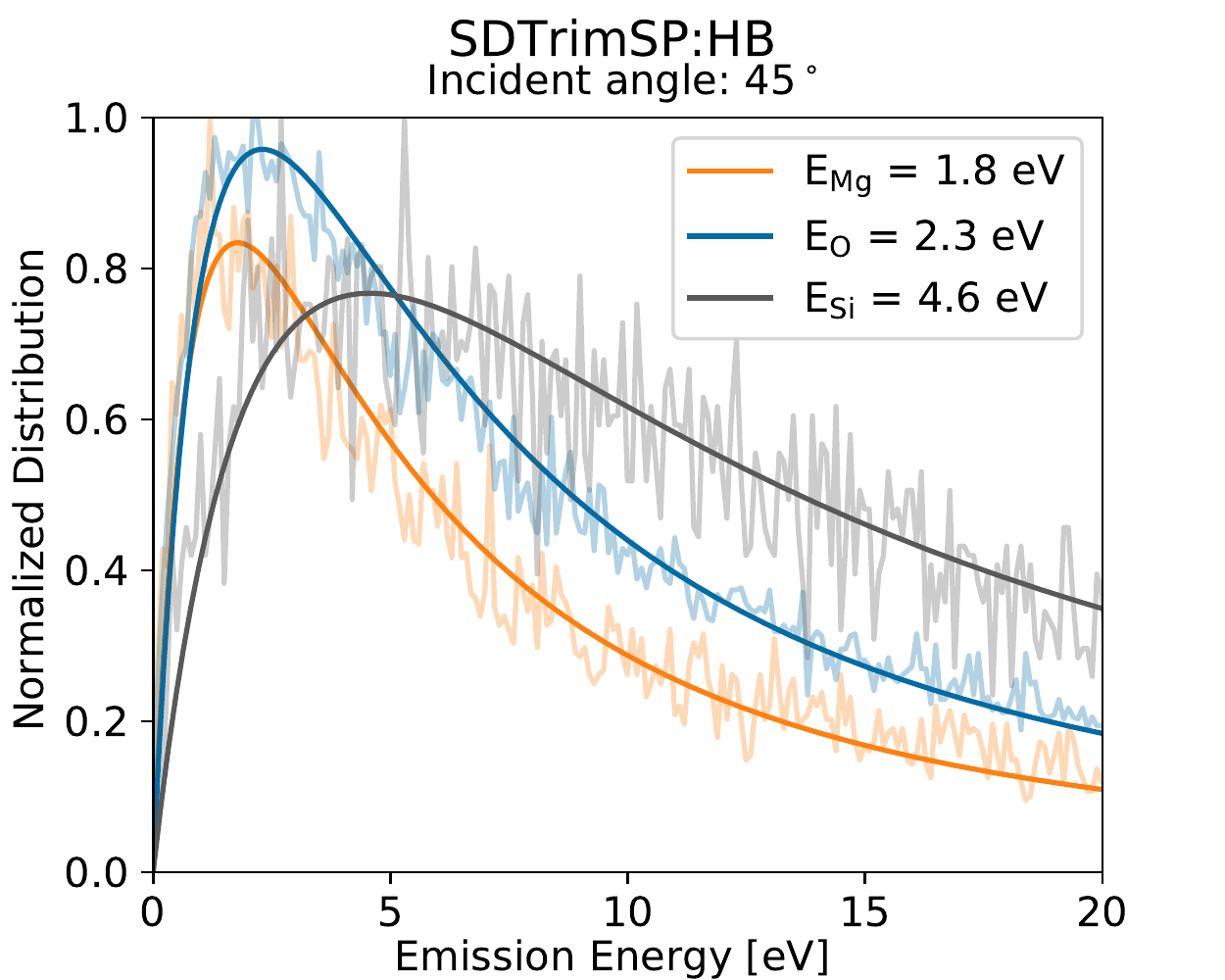}{0.49\textwidth}{}
    \fig{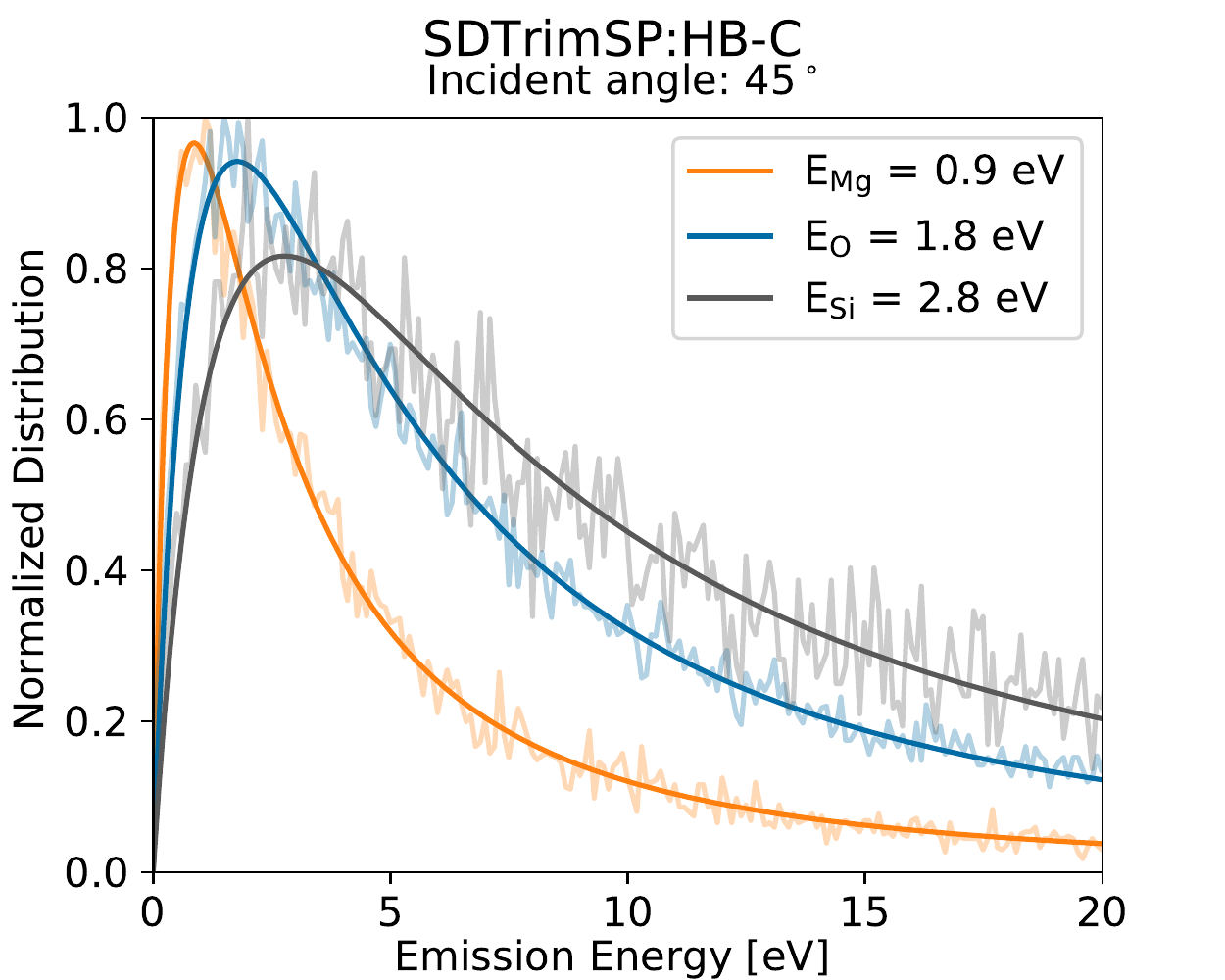}{0.49\textwidth}{}
    }
    \caption{Energy distributions of sputtered elements from enstatite irradiated with 4~keV~\ch{He+} at an angle of 45$^\circ$ based on different model assumptions. The energy in the legend corresponds to the peak energy of the Thompson fit function. Abbreviations: SBE -- tabulated enthalpy of sublimation as element surface binding energies; HB -- tabulated enthalpy of formation as bulk binding energy and enthalpy of sublimation as surface binding energies; C -- densities calculated based on compound densities and differentiation between compound-bound and un-bound atoms.
    \label{fig:edist_En}}
\end{figure*}

\section{Discussion} \label{sec:discussion}

\subsection{Sputter yield}\label{dis:yield}
We were able to confirm that it is of utmost importance to properly set the density of the irradiated sample. It is evident in Fig.~\ref{fig:WoEnVSexp} that under normal incidence, the HB-C model which recreates the mineral density adequately fits the experimental data best for both \ch{H+} and \ch{He+} irradiation results.

The experimental data of the H-irradiated wollastonite thin-film expresses a significant deviation from SDTrimSP predictions for the flat surface sputter behavior. This could so far not be explained \citep{Szabo2018}. Nevertheless, all the experimental data in Fig.~\ref{fig:WoEnVSexp} shows good agreement with the HB-C model close to normal incidence and up to at least 45$^\circ$. This is relevant for approximating irradiation of a realistic, rough surfaces, because yield enhancements between a flat and rough surface are generally small for incidence angles below 45$^\circ$ \citep[][]{Kustner1998,Biber2022}. This is not due to impacts realistically occurring at normal incidence in nature, but due to surface roughness leading to locally reduced incidence angles for shallow impinging ions and therefore flattened mass yield distributions. This is discussed in \cite{Biber2022} for enstatite irradiation experiments and was previously shown for rough Bo and Be surfaces \citep{Gauthier1990,Roth1991,Kustner1999}.

\subsection{Angular distribution}\label{res:angdist}
We observed, that no model can completely recreate the large polar tilt angle seen in experimental data (Fig.~\ref{fig:adist_En}). The model that comes closest is the HB model, which boasts large BBEs, subsequently leading to a rapid loss of energy with each recoil. The increased binding energy thus negatively affects the collision kinematics of long collision cascades and gives primary-knock-on collisions \citep[i.e., Fig.~2.6 in][]{Behrisch1991} a higher significance in the angular distribution of sputtered material. More random ejecta from long collision cascades which would lead to ejecta distributions close to normal is reduced. As a consequence, the tilt of the angular distribution increases. This behavior has also been observed on binary alloys, both experimentally and through MD simulations. There, atoms sputtered from the second atomic layer form angular distributions towards the surface normal whereas first-layer-emitted atoms have a broad distribution \cite{Schwebel1987,Whitaker1993,Gnaser1999}. In all but the HB and HB-C model, components with low BBEs (if any) exist at the irradiation equilibrium. Energy loss within the sample is therefore less significant, which reduces the contribution of first-layer-emitted atoms and causes a near circular plume of ejecta closer to the surface normal. 

The width of the angular distribution, quantified in the cosine fit exponents ($m$ and $n$, Fig~\ref{fig:adist_En}), is also tied to the surface binding energy. In all modeling approaches but the ones from \cite{Szabo2020a} and \cite{Hofsass2022} the used SBEs are identical and therefore the exponents are comparable. The BB model is most narrow (no surface potential, no refraction) and results in the lowest tilt angle with a visible forward-sputter contribution which is not able to significantly affect the tilt of the distribution. Both the HB-C and especially the HB model lead to a larger tilt due to preventing randomly distributed, low-energy particles to leave the surface and thus favoring forward-facing ejecta, which is observed as a peak around $-60^\circ$. Towards increasing incident angles relative to the surface normal ($>45^\circ$, not shown), the number of single knock-on recoils increases independent of the chosen model, enhancing the peak size of the forward-aligned ejecta. Local shallow incident angles are unlikely to contribute to sputtering of a realistic, rough and/or porous sample. This is motivated by the strong sputter yield decrease observed at shallow incidence, which is related to processes of shadowing and re-deposition \citep{Kustner1999,Cupak2021,Biber2022,Szabo2022c}. For this reason, the forward-facing peak at shallow incidence angles is not expected to be present for sputtering of regolith. Furthermore, the contribution to the total sputtered particles is negligible for non-shallow incident angles.

The sample roughness could in theory be another cause for the discrepancy between model and experimental data. The surface of the enstatite glassy thin-film was analyzed using an Atomic Force Microscope and its roughness was deemed negligible \citep{Biber2022}. Furthermore, when compared to the angular distribution of a rougher surface, the thin film angular distribution is nearly identical when normalized \citep[figures 2\&3 in ][]{Biber2022}. Roughness is therefore unlikely to account for the discrepancy seen in Figure~\ref{fig:adist_En}.

\subsection{Energy distribution}\label{dis:edist}
Energy distributions of particles from SB models follow Thompson distributions with peak energies close to 1/2 of the SBEs used. The HB model however reaches peak energies that are approximately equal to the SBEs used ($E_s(Mg)=1.5$, $E_s(O)=2.6$, and $E_s(Si)=4.7$) and the HB-C model shows elevated energies which are closer to SBE/2. At constant SBEs, the peaks of the energy distribution are widened with increasing bulk binding energies (Fig.~\ref{fig:edist_En}). Models which include a BBE experience a shift towards larger energies with a broadening of the energy distribution, as low-energy particles are not reflected back into the sample. This behavior follows the \ch{O2}-covered metal irradiation experiments performed by \cite{Dullni1984, Wucher1986} and \cite{Wucher1988}. Therefore the peak energies of the energy distributions, fitted by Thompson distributions, do not correspond to the enthalpy of sublimation $\Delta H_s$ of the atomic species but rather the combination of enthalpy of formation $\Delta H_f$ of the oxide present with $\Delta H_s$ \cite[Fig.~3 in][]{Dullni1984}. The expected energy distribution broadening in a system where \ch{O2} is present is thus recreated by both the HB and the HB-C~models with the same underlying assumptions, making it a valuable addition to the SB and BB models which, on the contrary, cannot. The results are  also  reminiscent of the broadening observed by increasing SBEs as in \cite{Morrissey2022}, and the conclusion is the same. Larger total binding energies lead to a larger high-energy fraction of the sputtered particles whilst reducing the number of ejected particles. In exospheres around solar-wind exposed surfaces, less abundant but more energetic particles would then be detectable farther from the surface.

\subsubsection{Inclusion of intermediary compounds}
It becomes evident from Figure~\ref{fig:edist_En}, that larger peak energies can be achieved if the atomic species remain in a bound condition. In the scope of this work we did not explore the formation of possible intermediates. The current implementation will always break up the compound and one of the products will continue to travel through the sample. If there are enough free elements available, only the original oxide can form, and therefore the model---for the example of \ch{SiO2}---is limited to:
\begin{equation}
    \ch{SiO2 <=> Si + O + O}
\end{equation}

A more sophisticated model would need to include the following reactions:
\begin{align}
  \ch{SiO2 &<=> Si + O2} \\
  \ch{SiO2 &<=> SiO + O} \\
  \ch{SiO &<=> Si + O} \\
  \ch{O2 &<=> O + O} \\
  \ch{Si2&<=> Si + Si}
\end{align}

which would reduce the number of un-bound atoms in the sample. The resulting energy distribution would thus lie closer to the hybrid model (HB) where atomic species are considered to remain bound in their compounds. To fully simulate the process of amorphization we would need to know what drives the stability of the different products within a mineral in irradiation equilibrium. 

\subsection{Effect of increased SBE}
To demonstrate the effect of an increased SBE, we compared the standard SB model and the newly implemented HB-C model with the results of \cite{Morrissey2022}. As of now, there are only SBEs available for Na in Na silicates with increasing coordination numbers (number of O atoms being a neighbor to Na). Therefore we only compare the results for albite \ch{NaAlSi3O8} irradiated by 1~keV \ch{H+} (Table~\ref{tab:morrissey}). For a static computation in SDTRimSP of albite with increased Na binding energies of $E_s(\text{Na})=7.9$~eV Morrissey reported a yield of $4.12\times10^{-4}$~Na$/$ion at normal incidence. If SDTrimSP is run in dynamic mode, the yield at the irradiation equilibrium is increased by a factor of two, to $7.90\times10^{-4}$~Na$/$ion. If compared to the yields of the SB model ($1.08\times10^{-3}$~Na$/$ion) and the HB-C model ($1.10\times10^{-3}$) the dynamic Na yields with $E_s(\text{Na})=7.9$~eV differ by 30\%. This similarity in SB and HB-C equilibrium yield is due to free Na atoms in the HB-C model behaving identical to the Na in the BB model. \ch{Na2O} having the lowest enthalpy of formation and therefore bound Na in the HB-C model is not prioritized in forming bonds with free O, causing an accumulation of Na in the surface layer at irradiation equilibrium as a result. The increase in density and BBE which is imbued in the HB-C model does therefore not apply to Na at the irradiation equilibrium as no surface \ch{Na2O} exists. The energy peak of the Morrissey approach ($E_s(\text{Na})=7.9$~eV) is, as expected, around 4~eV (approx. SBE/2~=~7.8/2) with the tilt angle exceeding the results of both the SB and the HB-C model by a factor of two and expressing a wide distribution as given by the large fit exponents ($m$ and $n$). In conclusion, the effect of increasing the SBE of Na is apparent not only in actual yields (-30\%) but also in the angular and energy energy distributions.

\begin{deluxetable}{lcccc}

    \tablecaption{Effect of an increased sodium surface binding energy on total yield and angular distribution from simulating 1~keV~\ch{H+} irradiation on albite (\ch{NaAlSi3O8})\label{tab:morrissey}}
    \tablehead{\colhead{} &\colhead{$E_s(\text{Na})$} & \colhead{$Y_\text{Na}$} & \colhead{$\phi_\text{tilt}(45^\circ)$} & \colhead{$m$, $n$} \\
    \colhead{} &\colhead{[eV]} & \colhead{[10$^{-3}\frac{\text{at}}{\text{ion}}$]} & \colhead{[$^\circ$]} & \colhead{[1]}}
        \startdata
        M22$^a$ & 7.9 & 0.41 & -    & -       \\ 
        SB               & 7.9 & 0.79 & 34.4 & 5.1, 1.5 \\
        SB               & 1.1 & 1.08 & 16.1 & 3.0, 2.0 \\
        HB-C            & 1.1 & 1.10 & 18.9 & 3.9, 2.3 \\
        \enddata
    \tablecomments{$^a$Computed in static mode; $Y_\text{Na}$ -- sodium sputter yield; $E_s(\text{Na})$ -- surface binding energy of sodium; $\phi_\text{tilt}(45^\circ)$ -- angular distribution tilt angle at an ion incidence angle of $45^\circ$ relative to surface normal; 
    $m$, $n$ -- cosine fit exponents}
    
    \tablerefs{M22: \cite{Morrissey2022}}
    
\end{deluxetable}

\subsection{Experiments needed for evaluation}
Both the angular and energy distribution data of sputtered minerals depend on the chosen surface and bulk binding energies. Extensive experiments to properly discriminate between different sputtered species as well as obtaining the species' energy distribution would be highly valuable for constraining surface and bulk binding energies. Obtaining energy distributions would give a needed insight on the energy peak broadening effect occurring on minerals. If this was available, further restrictions on realistic binding energies could be enforced whereas SBEs define the energy peak position and width and BBEs act as a `broadening agent' for further enhancing energy peak widths. As a side effect, the increasing and/or shifting of binding energies between SBE and BBE could achieve the desired forward tilt of the sputtered material whilst not degrading the agreement in total mass yields. 

It would be pleasing, although unlikely, if experimental data of energy and angular distributions could be recreated based on solely tabulated thermodynamic data. Nevertheless, we expect SBEs to be larger than tabulated, as demonstrated for an ideal, intact crystal lattice in MD by \cite{Morrissey2022}. Using one single SBE might not be appropriate to describe an altered sample however. SBEs at various degrees of alteration would be necessary to understand the evolution of the SBE with increasing level of amorphization. The correlation of SBE with coordination number shown by Morrissey is reminiscent of the SBE dependence on the degree of amorphization and a similar behavior is expected for the surfaces of irradiated samples \citep{Loeffler2009,Biber2022}. One should however refrain from adjusting the SBE like a fit parameter to best reproduce experimental data. For now we propose the use of the HB-C model for recreating experimental mass changes, with the enthalpy of sublimation as SBE and the enthalpy of formation of the mineral-forming compounds as BBE.

\section{Conclusions} \label{sec:conclusions}

We introduced a hybrid binding energy model in the binary collision approximation (BCA) code SDTrimSP with an underlying compound model which combines tabulated data for surface binding energies (SBE), bulk binding energies (BBE) as well as densities for mineral samples whilst differentiating between free and compound-bound components. In regards to previous modeling approaches we offer an alternative that minimizes the number of free parameters further and well reproduces experimental data. The new compound hybrid model (HB-C) merges the pure surface binding (SB) and bulk binding (BB) models while reproducing mineral properties. This includes proper mineral densities through tabulated compound data, but also combining surface and bulk binding energies, which leads to increased energy loss within the collision cascade, causing energy peak broadening as expected in a O-dominated system \citep[e.g.,][]{Dullni1984}. 

Although the differences between the SB and the HB-C model seem minor, the model infrastructure allows for further inclusions that are reasonable in terms of mineralogy and physics. Furthermore, comparisons with experimental sputter yields result in unprecedented agreement between $0^\circ$ (normal incidence) and $45^\circ$, a range which is especially of interest for modelers that require sputter yields as inputs. The HB-C model thus convinces on the following points: 1)~Good agreement with existing experimental data in parameter spaces relevant to exosphere modelers; 2)~Corrects for underestimation of the default sample density computation based on atomic densities by using tabulated densities of compounds instead; 3)~Sets surface binding energies and bulk binding energies based on tabulated enthalpy of sublimation and enthalpy of formation of compounds respectively, which allows for an universal application to minerals; 4)~Does not require setting parameters such as SBE, BBE, density, and cut-off energy (surface-binding-model four, isbv~=~4, in SDTrimSP), therefore greatly increasing the ease-of-use. For the time being, the HB-C model does an exemplary job in recreating experimental sputter data whilst producing reasonable energy and angular distributions of ejecta.

\begin{acknowledgements}
Financial support has been provided by the Swiss National Science Foundation Fund (200021L\_182771/1) as well as the Austrian Science Fund FWF (Project No. I 4101-N36) and by KKK\"O (Commission for the Coordination of Fusion research in Austria at the Austrian Academy of Sciences \"OAW). The authors gratefully acknowledge support from NASA's Solar System Exploration Research Virtual Institute (SSERVI)
via the LEADER team, Grant \#80NSSC20M0060.\\
\end{acknowledgements}

\software{SDTrimSP \citep{Mutzke2019}, TRIM (in SRIM package) \citep{Biersack1984,Ziegler2010}}

\clearpage

\clearpage

\appendix
\restartappendixnumbering


\section{Averaging the surface binding energies}\label{app:isbv2}
If we assume, like in \cite{Szabo2020a}, that the binding energy that has to be overcome is solely dependent on the number of bonds with O, called the  coordination number, the SBE of any component would be a function of the O content in the sample. A way to simulate this effect of the coordination number of atoms is to assume an averaged binding energy, which is a mass balance over all species present in the compound. In SDTrimSP, this is implemented as the surface-binding-model two \citep[isbv~=~2,][]{Mutzke2019}: 
\begin{equation}
    SBE = \sum{q_i Es_i},
\end{equation}
where $q_i$ is the concentration and $Es_i$ is the SBE of component $i$. This results in a single SBE for all components and therefore the compound. This was applied in \cite{Szabo2020a} in addition to the density correction to best fit wollastonite (\ch{CaSiO3}) data. To illustrate this effect, let us assume an increased $Es_O$ of 6.5~eV \citep{Szabo2020a} and compare it to the default $Es_O$ of 2.58247~eV. For nepheline, (\ch{NaAlSiO4}) this would result in an average $Es$ of 5.03~eV for all species instead of 2.79~eV with 

\begin{equation}
    \begin{aligned}
    q_{Na}&=q_{Al}=q_{Si}=1/7\\
    q_O&=4/7\\
    Es_{Na}&=1.11\,\mathrm{eV}\\
    Es_{Al}&=3.41\,\mathrm{eV}\\ 
    Es_{Si}&=4.66\,\mathrm{eV}\\
    Es_{O}&=2.58\,\mathrm{eV} \Rightarrow Es_{avg}=2.79\,\mathrm{eV}\\
    Es_{O}&=6.50\,\mathrm{eV} \Rightarrow Es_{avg}=5.03\,\mathrm{eV}
    \end{aligned}
\end{equation}

On first glance, this seems to work, as the suggested SBE for Na in a pristine, crystalline mineral is about 4.8~eV based on Molecular Dynamics (MD) simulations \citep{Morrissey2022}. In the case of the major rock forming mineral albite \citep[\ch{NaAlSi2O6}; $Es_{Na}=8.4$~eV][]{Morrissey2022}, the isbv~=~2 approximation with $Es_O=6.5$~eV nets an average SBE of 5.4~eV, which does not reproduce the high binding energies of Na suggested by MD. This suggests that adjusting SBEs based on a single component has its limits when it comes to simulating bond strengths of complex mineral structures. 


\clearpage
\newpage
\bibliography{references.bib}{}

\begin{thebibliography}{}
\expandafter\ifx\csname natexlab\endcsname\relax\def\natexlab#1{#1}\fi
\providecommand{\url}[1]{\href{#1}{#1}}
\providecommand{\dodoi}[1]{doi:~\href{http://doi.org/#1}{\nolinkurl{#1}}}
\providecommand{\doeprint}[1]{\href{http://ascl.net/#1}{\nolinkurl{http://ascl.net/#1}}}
\providecommand{\doarXiv}[1]{\href{https://arxiv.org/abs/#1}{\nolinkurl{https://arxiv.org/abs/#1}}}

\bibitem[{Arredondo {et~al.}(2019)Arredondo, Oberkofler, Schwarz-Selinger, von
  Toussaint, Burwitz, Mutzke, Vassallo, \& Pedroni}]{Arredondo2019}
Arredondo, R., Oberkofler, M., Schwarz-Selinger, T., {et~al.} 2019, Nuclear
  Materials and Energy, 18, 72, \dodoi{10.1016/J.NME.2018.12.007}

\bibitem[{Baker {et~al.}(2013)Baker, Poh, Odstrcil, Arge, Benna, Johnson,
  Korth, Gershman, Ho, McClintock, Cassidy, Merkel, Raines, Schriver, Slavin,
  Solomon, Tr{\'{a}}vn{\'{i}}{\v{c}}ek, Winslow, \& Zurbuchen}]{Baker2013}
Baker, D.~N., Poh, G., Odstrcil, D., {et~al.} 2013, Journal of Geophysical
  Research: Space Physics, 118, 45, \dodoi{10.1029/2012JA018064}

\bibitem[{Baretzky {et~al.}(1992)Baretzky, M{\"{o}}ller, \&
  Taglauer}]{Baretzky1992}
Baretzky, B., M{\"{o}}ller, W., \& Taglauer, E. 1992, Vacuum, 43, 1207,
  \dodoi{10.1016/0042-207X(92)90025-R}

\bibitem[{Behrisch \& Eckstein(2007)}]{Behrisch2007}
Behrisch, R., \& Eckstein, W. 2007, {Sputtering by particle bombardment :
  experiments and computer calculations from threshold to MeV energies}
  (Springer), 507

\bibitem[{Behrisch \& Wittmaack(1991)}]{Behrisch1991}
Behrisch, R., \& Wittmaack, K. 1991, {Sputtering by Particle Bombardment III:
  Characteristics of Sputtered Particles, Technical Applications} (Berlin,
  Heidelberg: Springer)

\bibitem[{Benkhoff {et~al.}(2010)Benkhoff, van Casteren, Hayakawa, Fujimoto,
  Laakso, Novara, Ferri, Middleton, \& Ziethe}]{Benkhoff2010}
Benkhoff, J., van Casteren, J., Hayakawa, H., {et~al.} 2010, Planetary and
  Space Science, 58, 2, \dodoi{10.1016/J.PSS.2009.09.020}

\bibitem[{Benninghoven {et~al.}(1987)Benninghoven, Rudenauer, \&
  Werner}]{Benninghoven1987}
Benninghoven, A., Rudenauer, F.~G., \& Werner, H.~W. 1987, {Secondary ion mass
  spectrometry: basic concepts, instrumental aspects, applications and trends}
  (John Wiley and Sons, New York, NY)

\bibitem[{Betz(1987)}]{Betz1987}
Betz, G. 1987, Nuclear Instruments and Methods in Physics Research Section B:
  Beam Interactions with Materials and Atoms, 27, 104,
  \dodoi{10.1016/0168-583X(87)90011-5}

\bibitem[{Betz \& Wien(1994)}]{Betz1994}
Betz, G., \& Wien, K. 1994, International Journal of Mass Spectrometry and Ion
  Processes, 140, 1, \dodoi{10.1016/0168-1176(94)04052-4}

\bibitem[{Biber {et~al.}(2022)Biber, Br{\"{o}}tzner, J{\"{a}}ggi, Szabo,
  Pichler, Cupak, Voith, Cserveny, Nenning, Mutzke, Moro, Primetzhofer, Mezger,
  Galli, Wurz, \& Aumayr}]{Biber2022}
Biber, H., Br{\"{o}}tzner, J., J{\"{a}}ggi, N., {et~al.} 2022, The Planetary
  Science Journal, 3, 271, \dodoi{10.3847/PSJ/ACA402}

\bibitem[{Biersack \& Haggmark(1980)}]{Biersack1980}
Biersack, J., \& Haggmark, L. 1980, Nuclear Instruments and Methods, 174, 257,
  \dodoi{10.1016/0029-554X(80)90440-1}

\bibitem[{Biersack \& Eckstein(1984)}]{Biersack1984}
Biersack, J.~P., \& Eckstein, W. 1984, Applied Physics A 1984 34:2, 34, 73,
  \dodoi{10.1007/BF00614759}

\bibitem[{Cassidy {et~al.}(2015)Cassidy, Merkel, Burger, Sarantos, Killen,
  McClintock, \& Vervack}]{Cassidy2015}
Cassidy, T.~A., Merkel, A.~W., Burger, M.~H., {et~al.} 2015, Icarus, 248, 547,
  \dodoi{10.1016/j.icarus.2014.10.037}

\bibitem[{Cupak {et~al.}(2021)Cupak, Szabo, Biber, Stadlmayr, Grave, Fellinger,
  Br{\"{o}}tzner, Wilhelm, M{\"{o}}ller, Mutzke, Moro, \& Aumayr}]{Cupak2021}
Cupak, C., Szabo, P.~S., Biber, H., {et~al.} 2021, Applied Surface Science,
  570, 151204, \dodoi{10.1016/J.APSUSC.2021.151204}

\bibitem[{Deer {et~al.}(1992)Deer, Howie, \& Zussman}]{Deer1992}
Deer, W.~A., Howie, R.~A., \& Zussman, J. 1992, {An introduction to the
  rock-forming minerals}, 2nd edn. (Harlow, Essex, England : New York, NY:
  Longman Scientific {\&} Technical), 696

\bibitem[{Domingue {et~al.}(2014)Domingue, Chapman, Killen, Zurbuchen, Gilbert,
  Sarantos, Benna, Slavin, Schriver, Tr{\'{a}}vn{\'{i}}{\v{c}}ek, Orlando,
  Sprague, Blewett, Gillis-Davis, Feldman, Lawrence, Ho, Ebel, Nittler, Vilas,
  Pieters, Solomon, Johnson, Winslow, Helbert, Peplowski, Weider, Mouawad,
  Izenberg, \& McClintock}]{Domingue2014}
Domingue, D.~L., Chapman, C.~R., Killen, R.~M., {et~al.} 2014, Space Science
  Reviews, 181, 121, \dodoi{10.1007/S11214-014-0039-5/FIGURES/20}

\bibitem[{Dukes \& Baragiola(2015)}]{Dukes2015}
Dukes, C., \& Baragiola, R. 2015, Icarus, 255, 51,
  \dodoi{10.1016/J.ICARUS.2014.11.032}

\bibitem[{Dukes {et~al.}(2011)Dukes, Chang, Fam{\'{a}}, \&
  Baragiola}]{Dukes2011}
Dukes, C.~A., Chang, W.~Y., Fam{\'{a}}, M., \& Baragiola, R.~A. 2011, Icarus,
  212, 463, \dodoi{10.1016/j.icarus.2011.01.027}

\bibitem[{Dullni(1984)}]{Dullni1984}
Dullni, E. 1984, Nuclear Instruments and Methods in Physics Research Section B:
  Beam Interactions with Materials and Atoms, 2, 610,
  \dodoi{10.1016/0168-583X(84)90276-3}

\bibitem[{Eckstein(1991)}]{Eckstein1991}
Eckstein, W. 1991, {Computer simulation of ion-solid interactions}, Vol.~10
  (Springer)

\bibitem[{Eckstein \& Preuss(2003)}]{Eckstein2003}
Eckstein, W., \& Preuss, R. 2003, Journal of Nuclear Materials, 320, 209,
  \dodoi{10.1016/S0022-3115(03)00192-2}

\bibitem[{Elphic {et~al.}(2014)Elphic, Delory, Hine, Mahaffy, Horanyi,
  Colaprete, Benna, \& Noble}]{Elphic2014}
Elphic, R.~C., Delory, G.~T., Hine, B.~P., {et~al.} 2014, {The Lunar Atmosphere
  and Dust Environment Explorer Mission},  Kluwer Academic Publishers,
  \dodoi{10.1007/s11214-014-0113-z}

\bibitem[{Fatemi {et~al.}(2020)Fatemi, Poppe, \& Barabash}]{Fatemi2020}
Fatemi, S., Poppe, A., \& Barabash, S. 2020, Journal of Geophysical Research:
  Space Physics, 125, e2019JA027706

\bibitem[{Gades \& Urbassek(1992)}]{Gades1992}
Gades, H., \& Urbassek, H.~M. 1992, Nuclear Instruments and Methods in Physics
  Research Section B: Beam Interactions with Materials and Atoms, 69, 232,
  \dodoi{10.1016/0168-583X(92)96012-N}

\bibitem[{Gamborino \& Wurz(2018)}]{Gamborino2018}
Gamborino, D., \& Wurz, P. 2018, Planetary and Space Science, 159, 97,
  \dodoi{10.1016/J.PSS.2018.04.021}

\bibitem[{Gauthier {et~al.}(1990)Gauthier, Eckstein, L{\'{a}}szl{\'{o}}, \&
  Roth}]{Gauthier1990}
Gauthier, E., Eckstein, W., L{\'{a}}szl{\'{o}}, J., \& Roth, J. 1990, Journal
  of Nuclear Materials, 176-177, 438, \dodoi{10.1016/0022-3115(90)90086-3}

\bibitem[{Gershman {et~al.}(2012)Gershman, Zurbuchen, Fisk, Gilbert, Raines,
  Anderson, Smith, Korth, \& Solomon}]{Gershman2012}
Gershman, D.~J., Zurbuchen, T.~H., Fisk, L.~A., {et~al.} 2012, Journal of
  Geophysical Research: Space Physics, 117, n/a, \dodoi{10.1029/2012JA017829}

\bibitem[{Glass {et~al.}(2022)Glass, Raines, Jia, Sun, Imber, Dewey, \&
  Slavin}]{Glass2022}
Glass, A.~N., Raines, J.~M., Jia, X., {et~al.} 2022, Journal of Geophysical
  Research: Space Physics, 127, e2022JA030969, \dodoi{10.1029/2022JA030969}

\bibitem[{Gnaser(1999)}]{Gnaser1999}
Gnaser, H. 1999, {Low-energy ion irradiation of solid surfaces}, Vol. 146
  (Berlin: Springer), 41--42

\bibitem[{Grava {et~al.}(2021)Grava, Killen, Benna, Berezhnoy, Halekas,
  Leblanc, Nishino, Plainaki, Raines, Sarantos, Teolis, Tucker, Vervack, \&
  Vorburger}]{Grava2021}
Grava, C., Killen, R.~M., Benna, M., {et~al.} 2021, Space Science Reviews 2021
  217:5, 217, 1, \dodoi{10.1007/S11214-021-00833-8}

\bibitem[{Hijazi {et~al.}(2017)Hijazi, Bannister, Meyer, Rouleau, \&
  Meyer}]{Hijazi2017}
Hijazi, H., Bannister, M.~E., Meyer, H.~M., Rouleau, C.~M., \& Meyer, F.~W.
  2017, Journal of Geophysical Research: Planets, 122, 1597,
  \dodoi{10.1002/2017JE005300}

\bibitem[{Hobler(2013)}]{Hobler2013}
Hobler, G. 2013, Nuclear Instruments and Methods in Physics Research Section B:
  Beam Interactions with Materials and Atoms, 303, 165,
  \dodoi{10.1016/J.NIMB.2012.11.022}

\bibitem[{Hofs{\"{a}}ss \& Stegmaier(2022)}]{Hofsass2022}
Hofs{\"{a}}ss, H., \& Stegmaier, A. 2022, Nuclear Instruments and Methods in
  Physics Research Section B: Beam Interactions with Materials and Atoms, 517,
  49, \dodoi{10.1016/J.NIMB.2022.02.012}

\bibitem[{Jackson(1975)}]{Jackson1975}
Jackson, D.~P. 1975, https://doi.org/10.1139/p75-194, 53, 1513,
  \dodoi{10.1139/P75-194}

\bibitem[{Janches {et~al.}(2021)Janches, Berezhnoy, Christou, Cremonese, Hirai,
  Hor{\'{a}}nyi, Jasinski, \& Sarantos}]{Janches2021}
Janches, D., Berezhnoy, A.~A., Christou, A.~A., {et~al.} 2021, Space Science
  Reviews, 217, 1

\bibitem[{Kazakov {et~al.}(2022)Kazakov, Milillo, Lazzarotto, Ivanovski,
  Mangano, Escalona-Mor{\'{a}}n, Mura, Massetti, Moroni, Orsini, Sordini,
  Angelis, Rispoli, Aronica, Vertolli, Alberti, Plainaki, Nuccilli, \&
  Noschese}]{Kazakov2022}
Kazakov, A., Milillo, A., Lazzarotto, F., {et~al.} 2022, in Europlanet Science
  Congress 2022 (Granada, Spain: Europla), \dodoi{10.5194/EPSC2022-810}

\bibitem[{{Ken Knight} \& Wehner(1967)}]{KenKnight1967}
{Ken Knight}, C.~E., \& Wehner, G.~K. 1967, {Investigation of sputtering
  effects on the moon's surface}, Tech. rep., National Aeronautics and Space
  Administration Headquarters Office of Lunar and Planetary Programs,
  Washington, D. C.

\bibitem[{Killen {et~al.}(2022)Killen, Morrissey, Burger, Vervack, Tucker, \&
  Savin}]{Killen2022}
Killen, R.~M., Morrissey, L.~S., Burger, M.~H., {et~al.} 2022, The Planetary
  Science Journal, 3, 139, \dodoi{10.3847/PSJ/AC67DE}

\bibitem[{K{\"{u}}stner {et~al.}(1998)K{\"{u}}stner, Eckstein, Dose, \&
  Roth}]{Kustner1998}
K{\"{u}}stner, M., Eckstein, W., Dose, V., \& Roth, J. 1998, Nuclear
  Instruments and Methods in Physics Research Section B: Beam Interactions with
  Materials and Atoms, 145, 320, \dodoi{10.1016/S0168-583X(98)00399-1}

\bibitem[{K{\"{u}}stner {et~al.}(1999)K{\"{u}}stner, Eckstein, Hechtl, \&
  Roth}]{Kustner1999}
K{\"{u}}stner, M., Eckstein, W., Hechtl, E., \& Roth, J. 1999, Journal of
  Nuclear Materials, 265, 22, \dodoi{10.1016/S0022-3115(98)00648-5}

\bibitem[{Lindhard \& Scharff(1961)}]{Lindhard1961}
Lindhard, J., \& Scharff, M. 1961, Physical Review, 124, 128,
  \dodoi{10.1103/PhysRev.124.128}

\bibitem[{Loeffler {et~al.}(2009)Loeffler, Dukes, \& Baragiola}]{Loeffler2009}
Loeffler, M.~J., Dukes, C.~A., \& Baragiola, R.~A. 2009, Journal of Geophysical
  Research, 114, E03003, \dodoi{10.1029/2008JE003249}

\bibitem[{Lue {et~al.}(2011)Lue, Futaana, Barabash, Wieser, Holmstrm, Bhardwaj,
  Dhanya, \& Wurz}]{Lue2011}
Lue, C., Futaana, Y., Barabash, S., {et~al.} 2011, Geophysical Research
  Letters, 38, 3202, \dodoi{10.1029/2010GL046215}

\bibitem[{Madey {et~al.}(2002)Madey, Johnson, \& Orlando}]{Madey2002}
Madey, T.~E., Johnson, R.~E., \& Orlando, T.~M. 2002, Surface Science, 500,
  838, \dodoi{10.1016/S0039-6028(01)01556-4}

\bibitem[{Mangano {et~al.}(2007)Mangano, Milillo, Mura, Orsini, {De Angelis},
  {Di Lellis}, \& Wurz}]{Mangano2007}
Mangano, V., Milillo, A., Mura, A., {et~al.} 2007, Planetary and Space Science,
  55, 1541, \dodoi{10.1016/J.PSS.2006.10.008}

\bibitem[{Martinez {et~al.}(2017)Martinez, Langlinay, Ponciano, da~Silveira,
  Palumbo, Strazzulla, Brucato, Hijazi, Agnihotri, Boduch, Cassimi, Domaracka,
  Ropars, \& Rothard}]{Martinez2017}
Martinez, R., Langlinay, T., Ponciano, C.~R., {et~al.} 2017, Nuclear
  Instruments and Methods in Physics Research Section B: Beam Interactions with
  Materials and Atoms, 406, 523, \dodoi{10.1016/J.NIMB.2017.01.042}

\bibitem[{{McNutt Jr} {et~al.}(2018){McNutt Jr}, Benkhoff, Fujimoto, \&
  Anderson}]{McNuttJr2018}
{McNutt Jr}, R.~L., Benkhoff, J., Fujimoto, M., \& Anderson, B.~J. 2018, in
  Mercury: The View after MESSENGER, ed. S.~C. Solomon, L.~R. Nittler, \& B.~J.
  Anderson (Cambridge, United Kingdom: Cambridge University Press), 544--569,
  \dodoi{10.1017/9781316650684}

\bibitem[{Milillo {et~al.}(2020)Milillo, Fujimoto, Murakami, Benkhoff, Zender,
  Aizawa, D{\'{o}}sa, Griton, Heyner, Ho, Imber, Jia, Karlsson, Killen,
  Laurenza, Lindsay, McKenna-Lawlor, Mura, Raines, Rothery, Andr{\'{e}},
  Baumjohann, Berezhnoy, Bourdin, Bunce, Califano, Deca, de~la Fuente, Dong,
  Grava, Fatemi, Henri, Ivanovski, Jackson, James, Kallio, Kasaba, Kilpua,
  Kobayashi, Langlais, Leblanc, Lhotka, Mangano, Martindale, Massetti, Masters,
  Morooka, Narita, Oliveira, Odstrcil, Orsini, Pelizzo, Plainaki, Plaschke,
  Sahraoui, Seki, Slavin, Vainio, Wurz, Barabash, Carr, Delcourt, Glassmeier,
  Grande, Hirahara, Huovelin, Korablev, Kojima, Lichtenegger, Livi, Matsuoka,
  Moissl, Moncuquet, Muinonen, Qu{\`{e}}merais, Saito, Yagitani, Yoshikawa, \&
  Wahlund}]{Milillo2020}
Milillo, A., Fujimoto, M., Murakami, G., {et~al.} 2020, Space Science Reviews,
  216, 1, \dodoi{10.1007/s11214-020-00712-8}

\bibitem[{M{\"{o}}ller \& Eckstein(1984)}]{Moller1984}
M{\"{o}}ller, W., \& Eckstein, W. 1984, Nuclear Instruments and Methods in
  Physics Research Section B: Beam Interactions with Materials and Atoms, 2,
  814, \dodoi{10.1016/0168-583X(84)90321-5}

\bibitem[{M{\"{o}}ller \& Posselt(2001)}]{Moller2001}
M{\"{o}}ller, W., \& Posselt, M. 2001, {TRIDYN{\_}FZR user manual, FZR Report
  No. 317}, Tech. rep., Forschungszentrum Rossendorf, Dresden, Germany

\bibitem[{Morrissey {et~al.}(2022)Morrissey, Tucker, Killen, Nakhla, \&
  Savin}]{Morrissey2022}
Morrissey, L.~S., Tucker, O.~J., Killen, R.~M., Nakhla, S., \& Savin, D.~W.
  2022, The Astrophysical Journal Letters, 925, L6,
  \dodoi{10.3847/2041-8213/AC42D8}

\bibitem[{Mura {et~al.}(2009)Mura, Wurz, Lichtenegger, Schleicher, Lammer,
  Delcourt, Milillo, Orsini, Massetti, \& Khodachenko}]{Mura2009}
Mura, A., Wurz, P., Lichtenegger, H. I.~M., {et~al.} 2009, Icarus, 200, 1,
  \dodoi{10.1016/J.ICARUS.2008.11.014}

\bibitem[{Mutzke {et~al.}(2019)Mutzke, Schneider, Eckstein, Dohmen, Schmid, von
  Toussaint, \& Badelow}]{Mutzke2019}
Mutzke, A., Schneider, R., Eckstein, W., {et~al.} 2019, {SDTrimSP Version
  6.00},  Max-Planck-Institut f{\"{u}}r Plasmaphysik

\bibitem[{N{\'e}non \& Poppe(2020)}]{Nenon2020}
N{\'e}non, Q., \& Poppe, A.~R. 2020, The Planetary Science Journal, 1, 69

\bibitem[{Orsini {et~al.}(2020)Orsini, Livi, Lichtenegger, Barabash, Milillo,
  Angelis, Phillips, \& Laky}]{Orsini2020}
Orsini, S., Livi, S., Lichtenegger, H., {et~al.} 2020, Space Science Reviews,
  217, 49

\bibitem[{Paige {et~al.}(2010)Paige, Foote, Greenhagen, Schofield, Calcutt,
  Vasavada, Preston, Taylor, Allen, \& Snook}]{Paige2010}
Paige, D.~A., Foote, M.~C., Greenhagen, B.~T., {et~al.} 2010, Space Science
  Reviews, 150, 125

\bibitem[{Pfleger {et~al.}(2015)Pfleger, Lichtenegger, Wurz, Lammer, Kallio,
  Alho, Mura, McKenna-Lawlor, \& Mart{\'{i}}n-Fern{\'{a}}ndez}]{Pfleger2015}
Pfleger, M., Lichtenegger, H., Wurz, P., {et~al.} 2015, Planetary and Space
  Science, 115, 90, \dodoi{10.1016/J.PSS.2015.04.016}

\bibitem[{Poppe {et~al.}(2018)Poppe, Farrell, \& Halekas}]{Poppe2018}
Poppe, A., Farrell, W., \& Halekas, J.~S. 2018, Journal of Geophysical
  Research: Planets, 123, 37

\bibitem[{Raines {et~al.}(2022)Raines, Dewey, Staudacher, Tracy, Bert,
  Sarantos, Gershman, Jasinski, Bowers, Fisher, \& Slavin}]{Raines2022}
Raines, J.~M., Dewey, R.~M., Staudacher, N.~M., {et~al.} 2022, Journal of
  Geophysical Research: Space Physics, e2022JA030397,
  \dodoi{10.1029/2022JA030397}

\bibitem[{Roth {et~al.}(1983)Roth, Bohdansky, \& Eckstein}]{Roth1983}
Roth, J., Bohdansky, J., \& Eckstein, W. 1983, Nuclear Instruments and Methods
  in Physics Research, 218, 751, \dodoi{10.1016/0167-5087(83)91077-3}

\bibitem[{Roth {et~al.}(1979)Roth, Bohdansky, \& Ottenberger}]{Roth1979}
Roth, J., Bohdansky, J., \& Ottenberger, W. 1979, {Data on low energy light ion
  sputtering},  Max-Planck-Institut f{\"{u}}r Plasmaphysik

\bibitem[{Roth {et~al.}(1991)Roth, Eckstein, Gauthier, \& Laszlo}]{Roth1991}
Roth, J., Eckstein, W., Gauthier, E., \& Laszlo, J. 1991, Journal of Nuclear
  Materials, 179-181, 34, \dodoi{10.1016/0022-3115(91)90010-5}

\bibitem[{Samartsev \& Wucher(2006)}]{Samartsev2006}
Samartsev, A.~V., \& Wucher, A. 2006, Applied Surface Science, 252, 6470,
  \dodoi{10.1016/J.APSUSC.2006.02.081}

\bibitem[{Schaible {et~al.}(2017)Schaible, Dukes, Hutcherson, Lee, Collier, \&
  Johnson}]{Schaible2017}
Schaible, M.~J., Dukes, C.~A., Hutcherson, A.~C., {et~al.} 2017, Journal of
  Geophysical Research: Planets, 122, 1968, \dodoi{10.1002/2017JE005359}

\bibitem[{Schaible {et~al.}(2020)Schaible, Sarantos, Anzures, Parman, \&
  Orlando}]{Schaible2020}
Schaible, M.~J., Sarantos, M., Anzures, B.~A., Parman, S.~W., \& Orlando, T.~M.
  2020, Journal of Geophysical Research: Planets, 125, e2020JE006479,
  \dodoi{10.1029/2020JE006479}

\bibitem[{Schwebel {et~al.}(1987)Schwebel, Pellet, \& Gautherin}]{Schwebel1987}
Schwebel, C., Pellet, C., \& Gautherin, G. 1987, Nuclear Instruments and
  Methods in Physics Research Section B: Beam Interactions with Materials and
  Atoms, 18, 525, \dodoi{10.1016/S0168-583X(86)80081-7}

\bibitem[{Sigmund(1969)}]{Sigmund1969}
Sigmund, P. 1969, Physical Review, 184, 383, \dodoi{10.1103/PhysRev.184.383}

\bibitem[{Solomon {et~al.}(2001)Solomon, McNutt, Gold, Acu{\~{n}}a, Baker,
  Boynton, Chapman, Cheng, Gloeckler, Head, Krimigis, McClintock, Murchie,
  Peale, Phillips, Robinson, Slavin, Smith, Strom, Trombka, \&
  Zuber}]{Solomon2001}
Solomon, S.~C., McNutt, R.~L., Gold, R.~E., {et~al.} 2001, Planetary and Space
  Science, 49, 1445, \dodoi{10.1016/S0032-0633(01)00085-X}

\bibitem[{Suzuki {et~al.}(2020)Suzuki, Yoshioka, Murakami, \&
  Yoshikawa}]{Suzuki2020}
Suzuki, Y., Yoshioka, K., Murakami, G., \& Yoshikawa, I. 2020, Journal of
  Geophysical Research: Planets, 125, e2020JE006472,
  \dodoi{10.1029/2020JE006472}

\bibitem[{Szabo {et~al.}(2022{\natexlab{a}})Szabo, Cupak, Biber, J{\"{a}}ggi,
  Galli, Wurz, \& Aumayr}]{Szabo2022c}
Szabo, P.~S., Cupak, C., Biber, H., {et~al.} 2022{\natexlab{a}}, Surfaces and
  Interfaces, 30, 101924, \dodoi{10.1016/J.SURFIN.2022.101924}

\bibitem[{Szabo {et~al.}(2018)Szabo, Chiba, Biber, Stadlmayr, Berger, Mayer,
  Mutzke, Doppler, Sauer, Appenroth, Fleig, Foelske-Schmitz, Hutter, Mezger,
  Lammer, Galli, Wurz, \& Aumayr}]{Szabo2018}
Szabo, P.~S., Chiba, R., Biber, H., {et~al.} 2018, Icarus, 314, 98,
  \dodoi{10.1016/J.ICARUS.2018.05.028}

\bibitem[{Szabo {et~al.}(2020{\natexlab{a}})Szabo, Biber, J{\"{a}}ggi, Brenner,
  Weichselbaum, Niggas, Stadlmayr, Primetzhofer, Nenning, Mutzke, Sauer, Fleig,
  Foelske-Schmitz, Mezger, Lammer, Galli, Wurz, \& Aumayr}]{Szabo2020a}
Szabo, P.~S., Biber, H., J{\"{a}}ggi, N., {et~al.} 2020{\natexlab{a}}, The
  Astrophysical Journal, 891, 100, \dodoi{10.3847/1538-4357/ab7008}

\bibitem[{Szabo {et~al.}(2020{\natexlab{b}})Szabo, Biber, J{\"{a}}ggi, Wappl,
  Stadlmayr, Primetzhofer, Nenning, Mutzke, Fleig, Mezger, Lammer, Galli, Wurz,
  \& Aumayr}]{Szabo2020b}
---. 2020{\natexlab{b}}, Journal of Geophysical Research: Planets, 125,
  e2020JE006583, \dodoi{10.1029/2020JE006583}

\bibitem[{Szabo {et~al.}(2022{\natexlab{b}})Szabo, Poppe, Biber, Mutzke,
  Pichler, J{\"{a}}ggi, Galli, Wurz, \& Aumayr}]{Szabo2022b}
Szabo, P.~S., Poppe, A.~R., Biber, H., {et~al.} 2022{\natexlab{b}}, Geophysical
  Research Letters, 49, e2022GL101232, \dodoi{10.1029/2022GL101232}

\bibitem[{Szymo{\'{n}}ski(1981)}]{Szymonski1981}
Szymo{\'{n}}ski, M. 1981, Physics Letters A, 82, 203,
  \dodoi{10.1016/0375-9601(81)90121-3}

\bibitem[{Thompson(1968)}]{Thompson1968}
Thompson, M.~W. 1968, PMag, 18, 377, \dodoi{10.1080/14786436808227358}

\bibitem[{{Van der Heide}(2014)}]{VanderHeide2014}
{Van der Heide}, P. 2014, {Secondary ion mass spectrometry: an introduction to
  principles and practices} (John Wiley {\&} Sons)

\bibitem[{Vorburger {et~al.}(2013)Vorburger, Wurz, Barabash, Wieser, Futaana,
  Lue, Holmstr{\"{o}}m, Bhardwaj, Dhanya, \& Asamura}]{Vorburger2013}
Vorburger, A., Wurz, P., Barabash, S., {et~al.} 2013, Journal of Geophysical
  Research: Space Physics, 118, 3937, \dodoi{10.1002/JGRA.50337}

\bibitem[{Whitaker {et~al.}(1993)Whitaker, Jones, Li, \& Watts}]{Whitaker1993}
Whitaker, T.~J., Jones, P.~L., Li, A., \& Watts, R.~O. 1993, Review of
  Scientific Instruments, 64, 452, \dodoi{10.1063/1.1144215}

\bibitem[{Winslow {et~al.}(2013)Winslow, Anderson, Johnson, Slavin, Korth,
  Purucker, Baker, \& Solomon}]{Winslow2013}
Winslow, R.~M., Anderson, B.~J., Johnson, C.~L., {et~al.} 2013, Journal of
  Geophysical Research: Space Physics, 118, 2213, \dodoi{10.1002/jgra.50237}

\bibitem[{Wucher \& Oechsner(1986)}]{Wucher1986}
Wucher, A., \& Oechsner, H. 1986, Nuclear Instruments and Methods in Physics
  Research Section B: Beam Interactions with Materials and Atoms, 18, 458,
  \dodoi{10.1016/S0168-583X(86)80071-4}

\bibitem[{Wucher \& Oechsner(1988)}]{Wucher1988}
---. 1988, Surface Science, 199, 567, \dodoi{10.1016/0039-6028(88)90921-1}

\bibitem[{Wurz(2005)}]{Wurz2005}
Wurz, P. 2005, in The Dynamic Sun: Challenges for Theory and Observations, Vol.
  600

\bibitem[{Wurz {et~al.}(2007)Wurz, Rohner, Whitby, Kolb, Lammer, Dobnikar, \&
  Mart{\'{i}}n-Fern{\'{a}}ndez}]{Wurz2007}
Wurz, P., Rohner, U., Whitby, J.~A., {et~al.} 2007, Icarus, 191, 486,
  \dodoi{10.1016/J.ICARUS.2007.04.034}

\bibitem[{Wurz {et~al.}(2010)Wurz, Whitby, Rohner,
  Mart{\'{i}}n-Fern{\'{a}}ndez, Lammer, \& Kolb}]{Wurz2010}
Wurz, P., Whitby, J.~A., Rohner, U., {et~al.} 2010, Planetary and Space
  Science, 58, 1599, \dodoi{10.1016/J.PSS.2010.08.003}

\bibitem[{Wurz {et~al.}(2022)Wurz, Fatemi, Galli, Halekas, Harada, J{\"{a}}ggi,
  Jasinski, Lammer, Lindsay, Nishino, Orlando, Raines, Scherf, Slavin,
  Vorburger, \& Winslow}]{Wurz2022}
Wurz, P., Fatemi, S., Galli, A., {et~al.} 2022, Space Science Reviews 2022
  218:3, 218, 1, \dodoi{10.1007/S11214-022-00875-6}

\bibitem[{Yamamura {et~al.}(1983)Yamamura, Itikawa, \& Itoh}]{Yamamura1983}
Yamamura, Y., Itikawa, Y., \& Itoh, N. 1983, {Angular dependence of sputtering
  yields of monatomic solids}, Tech. rep., Institute of Plasma Physics, Nagoya
  University, Nagoya

\bibitem[{Ziegler \& Biersack(1985)}]{Ziegler1985}
Ziegler, J.~F., \& Biersack, J.~P. 1985, Treatise on Heavy-Ion Science, 93,
  \dodoi{10.1007/978-1-4615-8103-1_3}

\bibitem[{Ziegler {et~al.}(2010)Ziegler, Ziegler, \& Biersack}]{Ziegler2010}
Ziegler, J.~F., Ziegler, M.~D., \& Biersack, J.~P. 2010, Nuclear Instruments
  and Methods in Physics Research, Section B: Beam Interactions with Materials
  and Atoms, 268, 1818, \dodoi{10.1016/j.nimb.2010.02.091}

\end{thebibliography}
\bibliographystyle{aasjournal}



\end{document}